\documentclass[reprint,amsmath,amssymb,aps]{revtex4-2}

\usepackage{graphicx}
\usepackage{dcolumn}
\usepackage{bm}
\usepackage[caption=false]{subfig} 
\usepackage{ragged2e}
\usepackage{svg}
\usepackage[colorlinks=true, linkcolor=blue, citecolor=blue, urlcolor=blue]{hyperref}
\usepackage{overpic}
\usepackage{float}
\usepackage{booktabs}
\usepackage[resetlabels]{multibib}

\begin{document}
\preprint{APS/123-QED}
\title{Comparative Optical Spectroscopy of Er$^{3+}$, Tm$^{3+}$, and Eu$^{3+}$ Molecular Complexes at Cryogenic Temperature}

\author{Farhad Rasekh$^{1,2}$}
\email{farhad.rasekh@ucalgary.ca}
\author{Sahil$^4$}
\author{Arsalan Mansourzadeh$^{1,2}$}
\author{Diogo Alves Gálico$^4$}
\author{David Hunger$^{5,6}$}
\author{Muralee Murugesu$^3$}
\author{Daniel Oblak$^{1,2}$}
\email{doblak@ucalgary.ca}
\affiliation{Institute for Quantum Science and Technology, University of Calgary, Calgary, AB, Canada$^1$,}
\affiliation{Department of Physics and Astronomy, University of Calgary, Calgary, AB, Canada$^2$,}
\affiliation{Department of Chemistry and Biomolecular Sciences, University of Ottawa, 10 Marie Curie, Ottawa, Ontario K1N 6N5, Canada$^3$,}
\affiliation{Department of Chemistry, University of Alberta, Alberta, Canada $^4$,}
\affiliation{Institute for Quantum Materials and technologies (IQMT), Karlsruhe
Institute of Technology, Karlsruhe, 76131,Germany$^5$,}
\affiliation{Physics Institute (PHI), Karlsruhe Institute of Technology, Karlsruhe, 76131,Germany$^6$.}

\begin{abstract}

Rare-earth molecular complexes are attracting increasing interest for quantum information applications because of their structural flexibility and chemical tunability. Here, we compare the cryogenic optical properties of Er$^{3+}$, Tm$^{3+}$, and Eu$^{3+}$ complexes in the same ligand environment. Optical spectroscopy shows differing behavior across the three systems: spectral hole burning is observed in the Eu$^{3+}$ complex, whereas no measurable hole is detected in the Er$^{3+}$ and Tm$^{3+}$ complexes. We measure laser-induced local heating and rethermalization in the Er$^{3+}$ and Tm$^{3+}$ powder-like samples, observing strong local heating and slow thermal relaxation in the cold-plate cryostat. The recovery occurs over several seconds and, for some Er$^{3+}$ excitation conditions, extends over tens of seconds. The comparison across the three ions nevertheless suggests that heating alone is unlikely to account for the full contrast. Rather, the different behaviors of Eu$^{3+}$, Tm$^{3+}$, and Er$^{3+}$ point to ion-dependent relaxation and decoherence processes associated with their electronic level structure and coupling to the molecular environment. Temperature-dependent linewidth and line-shift measurements are further consistent with phonon-mediated broadening through coupling to crystal-field levels.
\end{abstract}

\maketitle

\section{Introduction}

Advances in quantum communication and other emerging quantum technologies rely on material platforms capable of reliably storing and manipulating quantum states~\cite{lvovsky2009optical, heshami2016quantum, tittel2025quantum}. Rare-earth ions (REIs) are of particular interest for these applications because their electronic structure naturally suppresses environmental perturbations~\cite{stevenson2022erbium, goldner2015rare}. The partially filled 4f shells of optically active REIs are strongly shielded by the outer 5s and 5p orbitals, which limits interactions with the surrounding medium. As a result, when REIs are embedded in solid-state or molecular hosts, the essential characteristics of the 4f manifold remain largely preserved~\cite{liu2006spectroscopic, goldner2015rare}.
  
To harness these properties, REIs must be incorporated into a host material whose local structure and dynamics affect their optical and spin properties. Most prior work has focused on inorganic crystalline hosts~\cite{le2021twenty,wang2025nuclear,zhong2015nanophotonic,kolesov2012optical} and amorphous materials~\cite{saglamyurek2015quantum, rasekh2026ultra, bornadel2025hole}, which have enabled a broad range of demonstrations relevant to quantum technologies. More recently, molecular complexes have emerged as a promising alternative, with several studies reported over the past five years~\cite{bayliss2020optically, kumar2021optical,serrano2022ultranarrow, vasilenko2026optically}.   
 
Molecular complexes offer several unique advantages for REI--based quantum technologies. Their chemically identical coordination environments allow the synthesis of ensembles in which all REI sites are structurally uniform, reducing inhomogeneous broadening \cite{kumar2021optical,serrano2022ultranarrow}. Molecular design also enables \AA ngstrom-level control over ion--ion spacing, including the realization of binuclear complexes with fixed intramolecular separations and controlled ion--ion interactions \cite{kumar2021optical,vasilenko2026controlled}. It also permits the intentional incorporation of different REI elements within a common ligand framework \cite{bayliss2020optically,kumar2021optical,serrano2022ultranarrow}. Molecular complexes provide well-defined REI positions and coordination environments---an attractive feature shared with certain conventional crystalline hosts, such as $\mathrm{Cs}_2\mathrm{NaYCl}_6$ and $\mathrm{EuCl}_3\cdot6\mathrm{H}_2\mathrm{O}$~\cite{schwartz1974electron,ahlefeldt2013optical}.

Most lanthanide molecular complex studies to date have focused on europium (Eu$^{3+}$) \cite{kumar2021optical,serrano2022ultranarrow,vasilenko2026optically, kuppusamy2023observation}. Here, we instead use the same molecular host to compare three different REI ions, thulium (Tm$^{3+}$), erbium (Er$^{3+}$), and europium (Eu$^{3+}$), in
\[
    [\mathrm{Ln}(acac)_3(phen)],
\]
where $\mathrm{Ln} = \mathrm{Er}^{3+}$, $\mathrm{Tm}^{3+}$, and $\mathrm{Eu}^{3+}$, acac denotes acetylacetonate, and phen denotes 1,10-phenanthroline. This comparison is useful because the three ions bring distinct spectroscopic characteristics to an otherwise similar chemical environment. Eu$^{3+}$ serves as an important molecular benchmark because coherent optical effects such as spectral hole burning have already been observed in Eu$^{3+}$-based complexes \cite{kumar2021optical,serrano2022ultranarrow, kuppusamy2023observation}. Er$^{3+}$ is attractive because its optical transition near $1.5\,\mu\mathrm{m}$ lies in the telecom band~\cite{saglamyurek2015quantum, saglamyurek2016multiplexed,jin2015telecom}, while Tm$^{3+}$ provides a complementary non-Kramers case with a simpler low-energy level structure than Er$^{3+}$~\cite{saglamyurek2011broadband, davidson2020improved}.

Different spectral-hole-burning (SHB) behavior is expected from the electronic and isotopic structures of the three ions. Eu$^{3+}$ is a non-Kramers ion with a $J=0$ ground state, suppressing first-order electronic magnetic sensitivity and direct electronic spin--phonon relaxation; both naturally abundant Eu isotopes also have nuclear spin, allowing all Eu$^{3+}$ ions, in principle, to participate in nuclear-spin shelving \cite{konz2003temperature,kuppusamy2023observation}. Er$^{3+}$ is instead a Kramers ion with a strong electronic magnetic moment, making holes stored in electronic Zeeman levels more susceptible to magnetic noise and rapid relaxation. Moreover, the Er$^{3+}$ complexe studied here have natural isotopic abundance, for which only approximately 23\% of the ions are $^{167}$Er with nonzero nuclear spin \cite{guillot2006hyperfine}. Tm$^{3+}$ is non-Kramers but has a $J=6$ ground-state manifold and therefore lacks the $J=0$ protection of Eu$^{3+}$ against coupling to the environment.

Against this background, the central question of this work is how the expected ion-dependent properties and the experimental thermal environment together shape coherence-related behavior within the same molecular host. To address this question, we measure the absorption spectra of the Er$^{3+}$, Tm$^{3+}$, and Eu$^{3+}$ complexes at room and cryogenic temperatures and determine the optical lifetime of the Eu$^{3+}$ complex. We then perform SHB experiments \cite{macfarlane1987spectral, konz2003temperature, hastings2008spectral} and observe persistent spectral holes only in the Eu$^{3+}$ complex, while no measurable hole is detected in the Er$^{3+}$ and Tm$^{3+}$ complexes under the explored conditions. Because the Eu$^{3+}$ experiments were performed in a liquid-helium bath with substantially greater cooling power than the closed-cycle cold-plate cryostat used for the Er$^{3+}$ and Tm$^{3+}$ measurements, this comparison does not immediately distinguish intrinsic ion-dependent behavior from differences in thermal environment. We therefore examine laser-induced local heating and slow rethermalization in the Er$^{3+}$ and Tm$^{3+}$ powder samples, and finally study the temperature dependence of the optical absorption linewidth and transition energy. Together, these measurements allow us to compare how thermal conditions and the different electronic structures of Eu$^{3+}$, Tm$^{3+}$, and Er$^{3+}$ shape the optical coherence properties of these rare-earth molecular complexes.

\begin{figure*}[t]
    \centering
    \subfloat[]{%
        \includegraphics[width=0.32\textwidth]{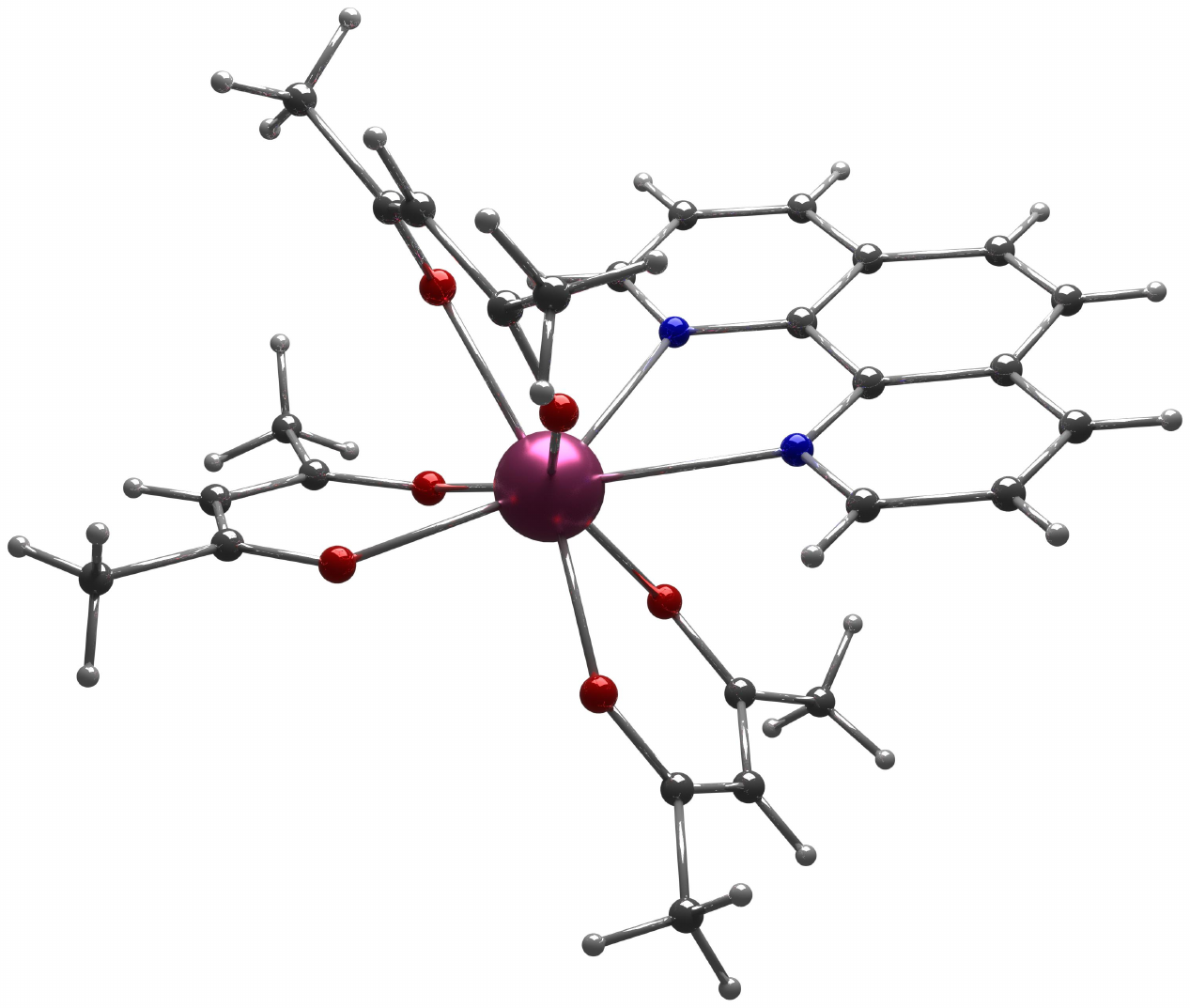}%
        \label{fig:molecule}
    }
    \subfloat[]{%
        \includegraphics[width=0.32\textwidth]{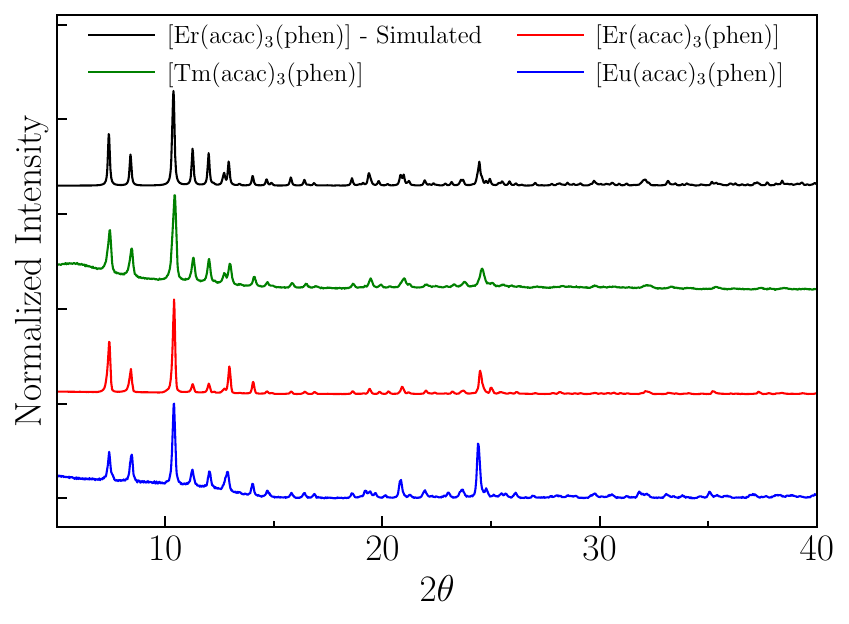}%
        \label{fig:PXRD}
    }
    \subfloat[]{%
        \includegraphics[width=0.32\textwidth]{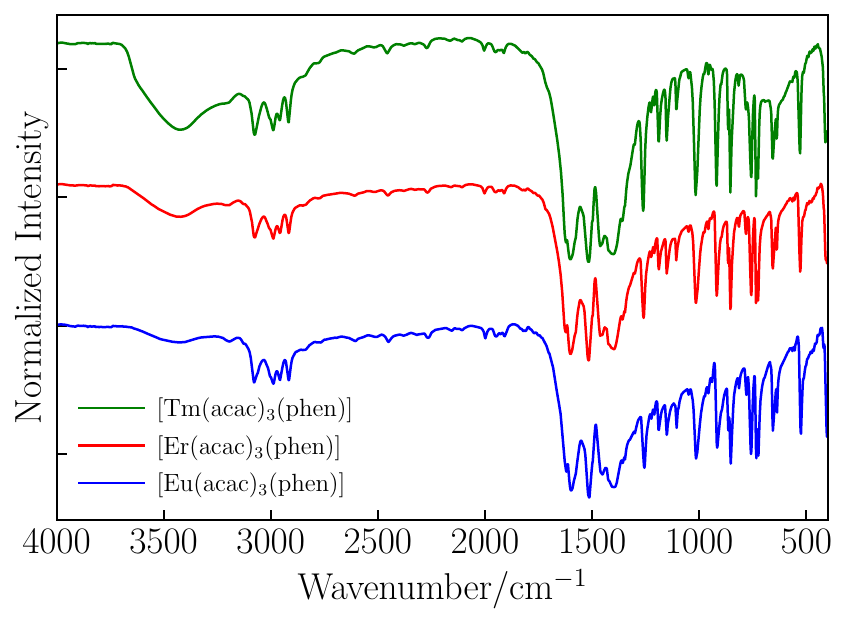}%
        \label{fig:FTIR}
    } 
    \caption{(a) Crystal structure of $[\mathrm{Er}(acac)_3(\mathrm{phen})]$ based on CCDC deposition 638310~\cite{neelgund2007tris}. Color code: Er, purple; O, red; N, blue; C, dark gray; H, light gray. (b) Experimental PXRD patterns for $[\mathrm{Ln}(acac)_3(\mathrm{phen})]$ complexes and simulated pattern for the $[\mathrm{Er}(acac)_3(\mathrm{phen})]$ complex. (c) FTIR spectra for $[\mathrm{Ln}(acac)_3(\mathrm{phen})]$ complexes. Throughout this work, the Er$^{3+}$, Tm$^{3+}$, and Eu$^{3+}$ data are represented by red, green, and blue color families, respectively, with different shades used to distinguish measurement conditions or datasets.}
    \label{fig:Chemistry}
\end{figure*}

\section{Methods }

Two distinct experimental configurations were used in this work: one for the Er$^{3+}$ and Tm$^{3+}$ complexes and another for the Eu$^{3+}$ complex. The use of two different experimental setups was dictated by the fact that the measurements were carried out in two separate laboratories, each with its own available cryogenic and optical infrastructure. Specifically, the Er$^{3+}$ and Tm$^{3+}$ measurements were performed in a pumped-$^3$He cryostat with free-space optical access, whereas the Eu$^{3+}$ measurements were conducted in a liquid-helium bath cryostat without free-space access, necessitating a fiber-coupled geometry.

The powder samples consisted of ensembles of as-grown, needle-like microcrystallites, with lengths ranging from a few micrometers to the millimeter scale and cross-sectional dimensions spanning from several hundred nanometers to a few micrometers. These crystallites were used without mechanical grinding or crushing. Throughout this work, the term powder refers to the particulate, powder-like appearance of these ensembles, rather than to a material produced by pulverizing larger crystals.
For the Er$^{3+}$ and Tm$^{3+}$ measurements, the powder was loaded into a custom-designed sample holder (Figure S1a in supporting information), which had a front aperture of approximately 700~$\mu$m, a back aperture of about 3~mm, and a pocket depth of roughly 400~$\mu$m. The sample was confined between two glass windows.

The holder was mounted in a continuously pumped-$^3$He Bluefors cryostat operated at approximately 500~mK and equipped with free-space optical access. A laser beam from an external-cavity diode laser (ECDL), with a diameter of approximately 250~$\mu$m, was focused onto the sample. The transmitted or emitted light was collected by a high-numerical-aperture lens positioned directly behind the sample inside the cryostat and subsequently focused onto a photodetector~(New Focus 2053). Depending on the measurement protocol, the laser was modulated using an acousto-optic modulator (AOM, Brimrose GPM-200-50-1300/1550), an optical switch, and a phase modulator. Separate lasers were used for the Tm$^{3+}$ and Er$^{3+}$ complexes, namely a MOGLabs continuous-wave (cw) laser for Tm$^{3+}$ and a Toptica DL Pro cw laser for Er$^{3+}$, corresponding to their optical transitions at 793~nm and 1530~nm, respectively. 

For the Er$^{3+}$ and Tm$^{3+}$ SHB measurements, the burn--wait--read pulse sequence was generated using the AOM and an electro-optic phase modulator (EOM). The EOM was driven with a serrodyne waveform to transfer the optical power predominantly into a single frequency-shifted sideband. During the read pulse, whose duration was varied from several microseconds to several milliseconds, the serrodyne frequency was swept from tens of MHz to a few GHz so that the shifted sideband scanned through zero detuning relative to the burn frequency. 

\begin{figure*}[!t]
  \centering

  \subfloat[]{%
    \includegraphics[width=0.32\textwidth]{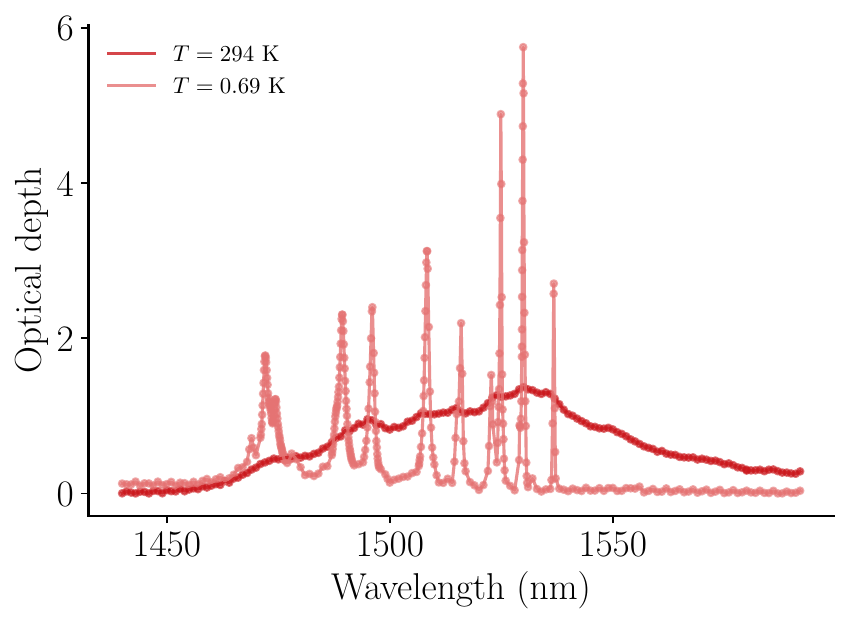}%
    \label{fig:OD-Er-CT}
  }
  \subfloat[]{%
    \includegraphics[width=0.32\textwidth]{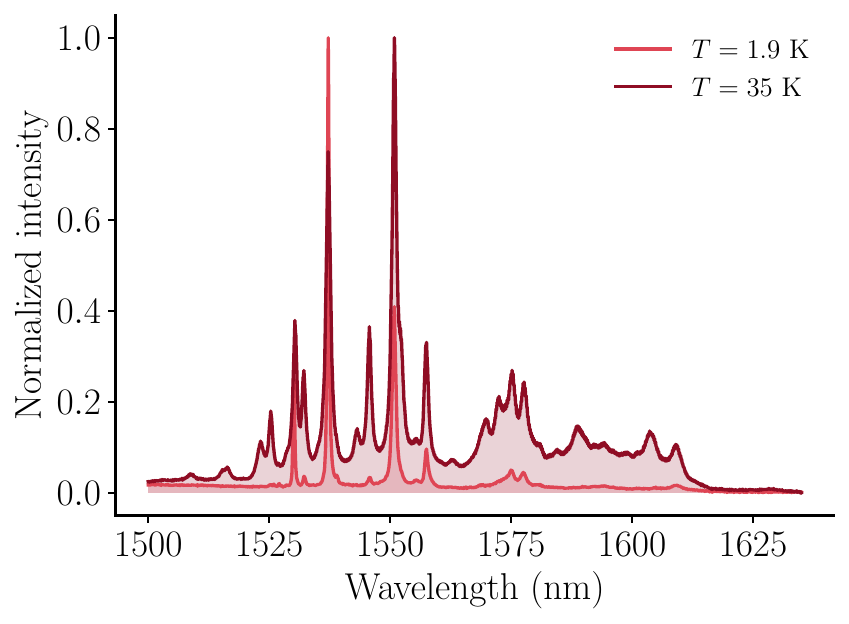}%
    \label{fig:PLE}
  }
    \subfloat[]{%
    \includegraphics[width=0.32\textwidth]{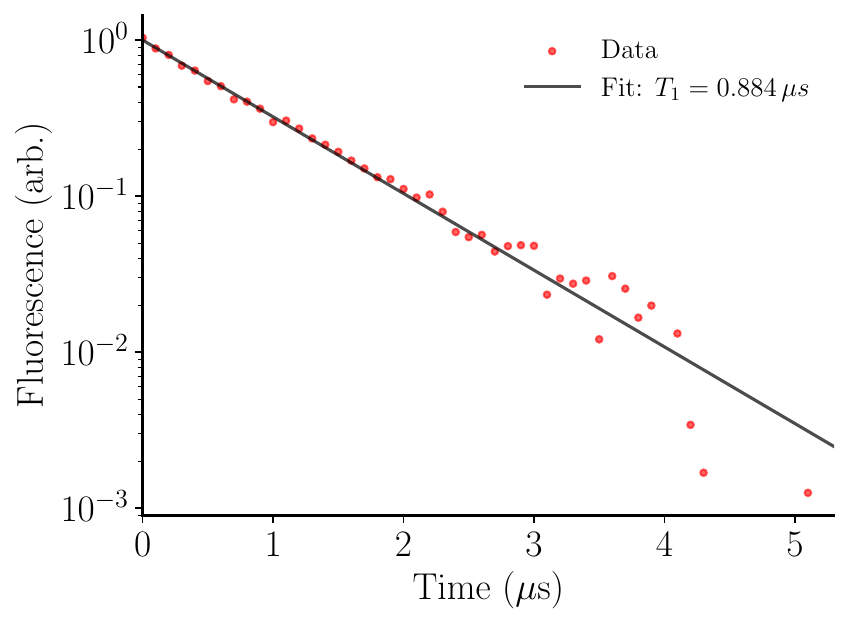}%
    \label{fig:Er-lifetime}
  }

  \subfloat[]{%
    \includegraphics[width=0.32\textwidth]{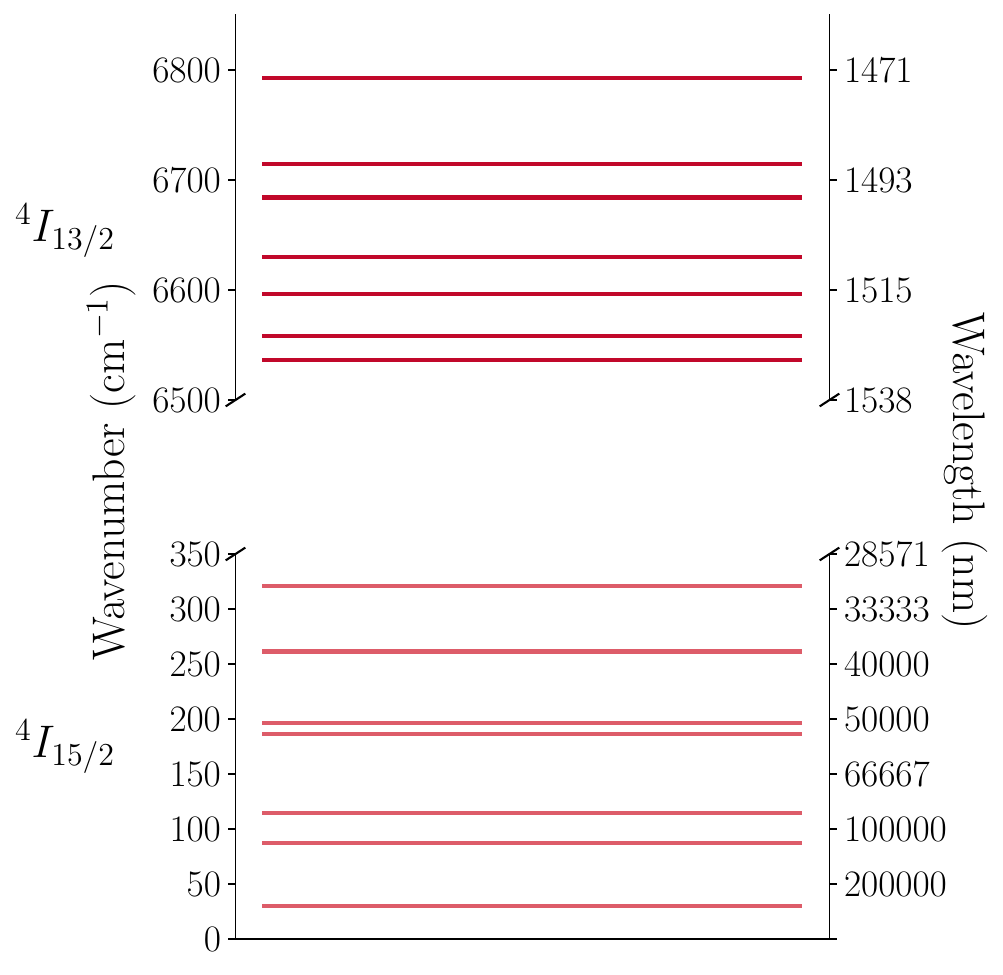}%
    \label{fig:Energy-levels-Er}
  }
    \subfloat[]{%
    \includegraphics[width=0.32\textwidth]{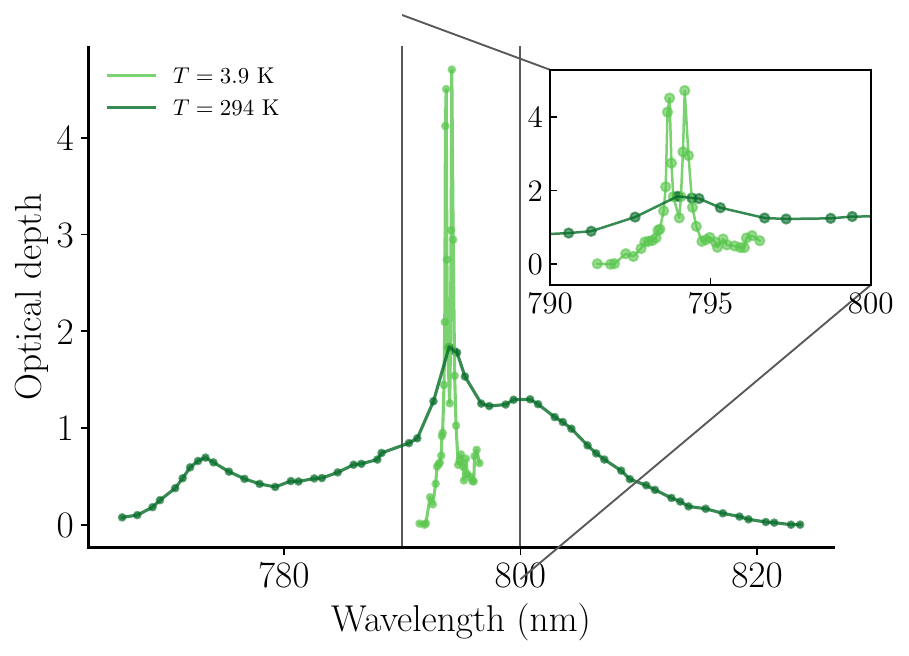}%
    \label{fig:OD-Tm-RT}
  }

  \caption{
    \justifying
    (a) Absorption profile of Er$^{3+}$ ions in the 100\%-concentration Er molecular complex measured at room temperature (dark red) and at 688~mK (light red). Seven distinct absorption peaks are observed, corresponding to the seven crystal-field levels of the excited-state manifold $^4I_{13/2}$. The lowest-energy peak has an inhomogeneous FWHM of $40.2 \pm 1.1$~GHz.
    (b) Photoluminescence spectra of Er$^{3+}$ ions in the Er complex measured at 1.9~K (light red) and 35~K (dark red) by exciting the ions to the $^4I_{11/2}$ level near $\sim$980~nm. 
    (c) Fluorescence intensity decay of Er$^{3+}$ ions in the Er$^{3+}$ complex. A single-exponential fit gives an excited-state lifetime of $0.884\pm0.013$ $\mu s$. 
    (d) Energy-level diagram of Er$^{3+}$ ions in the Er complex, showing eight crystal-field levels in the ground-state manifold $^4I_{15/2}$ and seven levels in the excited-state manifold $^4I_{13/2}$. 
    (e) Absorption profile of the $^3H_6 \rightarrow {}^3H_4$ transition of Tm$^{3+}$ ions in the 100\%-concentration Tm molecular complex measured at room temperature (dark green) and at 3.9~K (light green). The lowest-energy peak has an inhomogeneous FWHM of $64.4 \pm 2.0$~GHz.  
  }
  \label{fig:absorption}
\end{figure*}

The Er$^{3+}$ photoluminescence spectra shown in Figure~\ref{fig:PLE} were collected using single crystals mounted in an Oxford Instruments SpectromagPT cryostat. The sample was excited with a 980~nm laser diode, and the emitted light was collected using a set of lenses and directed to a Shamrock SR-750 spectrograph equipped with a 300~lines/mm grating blazed at 1700~nm and an Andor iDus InGaAs 490 detector.

For the Eu$^{3+}$-complex measurements, the sample was positioned between two multimode fibers with core diameters of 200~$\mu$m and 1000~$\mu$m, with the smaller fiber used for excitation and the larger one for collection. This assembly was immersed in a liquid-helium bath at 4.2~K. Optical excitation was provided by a Sirah Matisse~2DX dye laser in combination with an acousto-optic modulator (AOM, Gooch\&Housego 3200-121) operated in a double-pass configuration, enabling precise control over both the amplitude and frequency of the optical pulses. The optical signals were detected using a Thorlabs avalanche photodetector (APD, APD130A2/M). For the Eu$^{3+}$ SHB measurements, a single AOM was used both to generate the burn--wait--read pulse sequence and to sweep the optical frequency during the read pulse.

\section{Results}

In this section, we first establish the main optical transitions of the Er$^{3+}$, Tm$^{3+}$, and Eu$^{3+}$ complexes and, in the case of Er$^{3+}$, resolve the relevant crystal-field structure from absorption and photoluminescence measurements. We then compare their spectral-hole-burning behavior, which reveals a clear contrast between the Eu$^{3+}$ complex and the Er$^{3+}$ and Tm$^{3+}$ complexes. Because the Eu$^{3+}$ measurements were performed in a liquid-helium bath, whereas the Er$^{3+}$ and Tm$^{3+}$ measurements were carried out in a cold-plate cryostat with substantially less efficient thermalization, we next examine laser-induced local heating and rethermalization in the Er$^{3+}$ and Tm$^{3+}$ powder samples. Finally, we study the temperature dependence of the optical absorption linewidth and transition energy in Er$^{3+}$ and Tm$^{3+}$ in order to assess how thermal broadening processes evolve across these complexes.

\subsection{Synthesis and Structural Characterization}\label{Synthesis and Structural Characterization}
$[\mathrm{Ln}(acac)_3(phen)]$ complexes were prepared following a previously reported straightforward procedure ~\cite{galico2025magneto}. Equimolar amounts (1 mmol each) of $[\mathrm{Ln}(acac)_3(H_2O)_2]$ and 1,10-phenanthroline (phen) were dissolved in 20 mL of hot acetone and stirred for 10 min. The resulting solution was filtered and stored in a refrigerator ($\approx 4\,^{\circ}\mathrm{C}$). Needle-like single crystals formed within 20 min, and the products were isolated in yields of up to 80\% after 24 h. Powder X-ray diffraction (PXRD) patterns of the three complexes, collected on a Bruker D8 Advance diffractometer, match the expected structures confirming the three complexes are isostructural (Figure~\ref{fig:PXRD}).  The molecular structure of the Er$^{3+}$ complex is shown in Figure~\ref{fig:molecule}. In these complexes, the $\mathrm{Ln}^{3+}$ ion is located in a $\mathrm{O}_6\mathrm{N}_2$ chemical environment with a distorted square antiprismatic symmetry ($D_{4d}$ point group). FTIR spectra were recorded in a Thermo Fisher Nicolet 8700 spectrometer with the samples dispersed in KBr pellets (Figure~\ref{fig:FTIR}) and display the rich vibrational structure characteristic of these molecular complexes. Prominent high-energy bands appear in the 2915--3076 cm$^{-1}$ ($\sim 3431$--$3251~nm$) region, corresponding to alkyl and aromatic C--H stretching modes. These vibrations are known to strongly influence the optical properties of $\mathrm{Ln}^{3+}$ materials because they can promote efficient non-radiative deactivation through multiphonon relaxation. Consistent with strong nonradiative quenching, the measured lifetime of the Er$^{3+}$ $^4I_{13/2}$ excited state is only $T_1 = 0.884 \pm 0.013~\mu$s (Figure~\ref{fig:Er-lifetime}), approximately three to four orders of magnitude shorter than the millisecond-scale lifetimes commonly reported for the same state in inorganic crystalline hosts \cite{bottger2003material,stevenson2022erbium}. The first overtones of these C--H stretches typically occur in the 5800--6200 cm$^{-1}$ ($\sim 1724$--$1613~nm$) range and can effectively bridge energy gaps of similar magnitude in many $\mathrm{Ln}^{3+}$ ions. Additionally, combination bands (C--H stretch + C--H bend) are observed around 4200--4500 cm$^{-1}$ ($\sim 2381$--$2222~nm$) ~\cite{teixeira2016thermal}.

\subsection{Optical Absorption Profile and PLE Spectrum}\label{Optical Absorption Profile and PLE Spectrum}
We begin by examining the optical absorption of the Er$^{3+}$, Tm$^{3+}$, and Eu$^{3+}$ complexes, which provide direct information about their transition energies, inhomogeneous broadening, and crystal-field structure. To quantify the absorption strength, we use the Beer--Lambert law \cite{swinehart1962beer},

\begin{equation}
I_{out} = I_{in} e^{-\alpha d} 
\label{eq:1}
\end{equation}
where $I_{in}$ and $I_{out}$ are the incident and transmitted light intensities, respectively, $d$ is the sample thickness, and $\alpha$ is the absorption coefficient. Using this framework, we first discuss the absorption measurements of the Er$^{3+}$ and Tm$^{3+}$ complexes, together with the photoluminescence (PL) spectra of the Er$^{3+}$ complex, followed by the optical characterization of the Eu$^{3+}$ complex.

\subsubsection{Er$^{3+}$ and Tm$^{3+}$}

\begin{figure*}[!t]
  \centering

  \subfloat[]{%
    \includegraphics[width=0.32\textwidth]{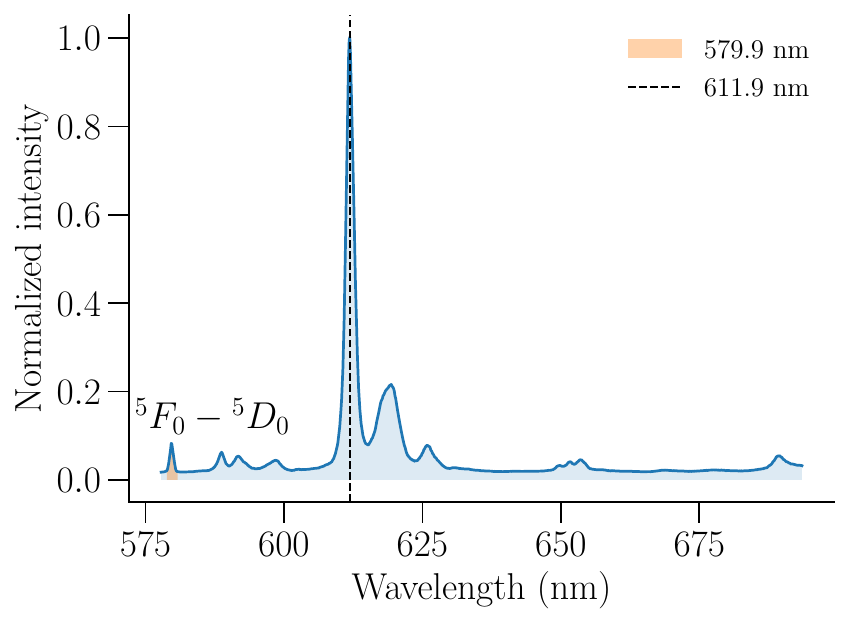}%
    \label{fig:Eu-spectrum}
  }\hfill
  \subfloat[]{%
    \includegraphics[width=0.32\textwidth]{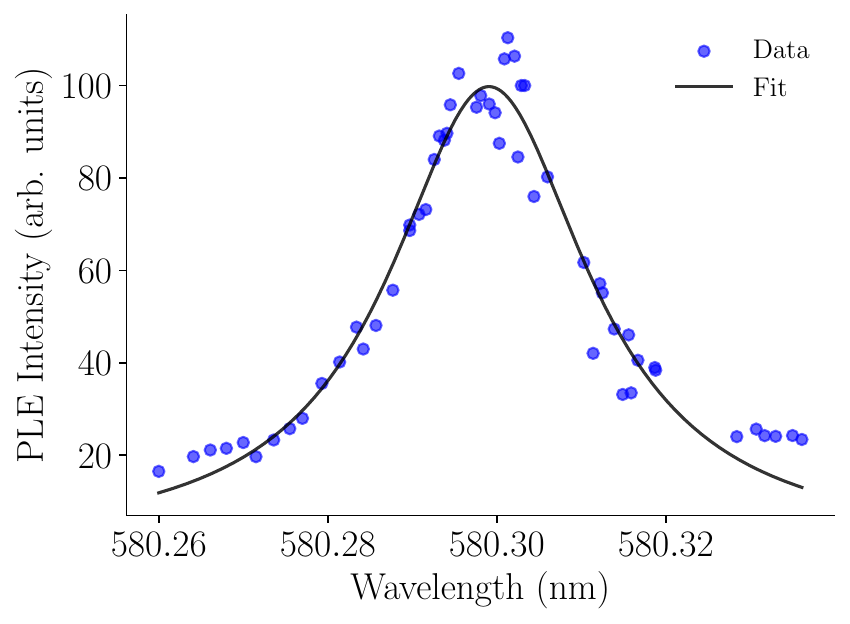}%
    \label{fig:Eu-PL}}
  \hfill
  \subfloat[]{%
    \includegraphics[width=0.32\textwidth]{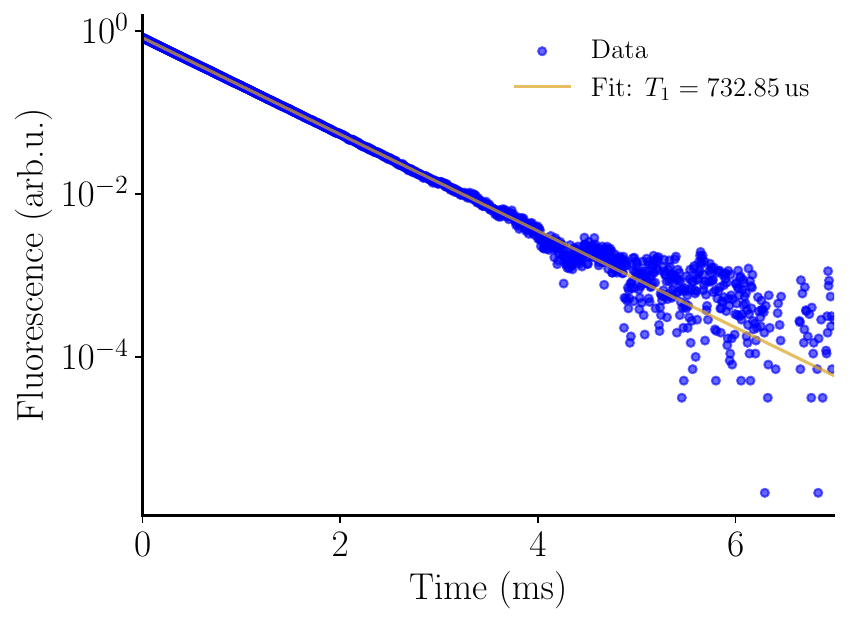}%
    \label{fig:Eu-T1}
  }

  \subfloat[]{%
    \includegraphics[width=0.32\textwidth]{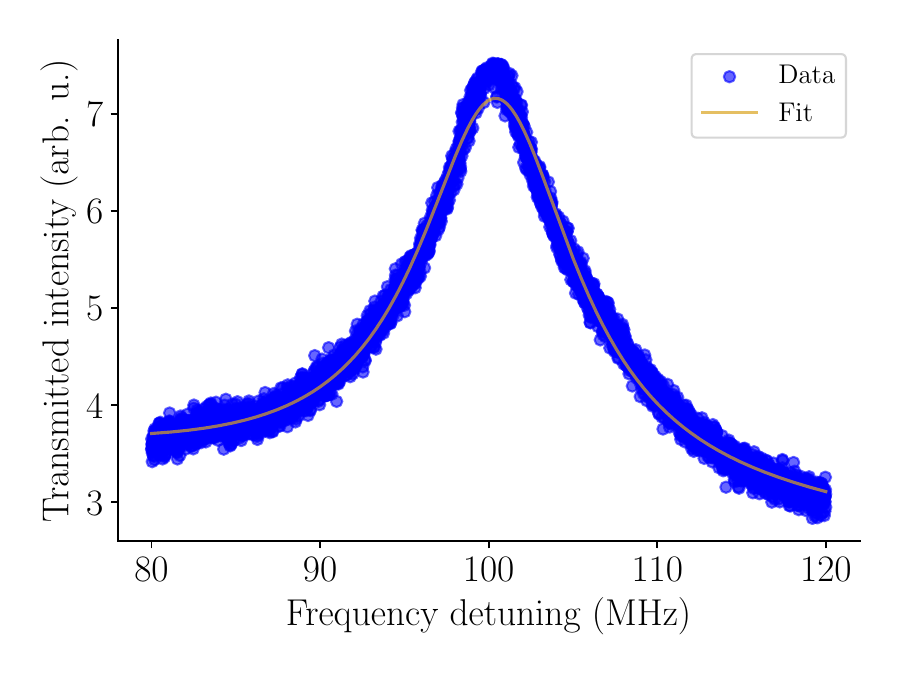}%
    \label{fig:Eu-SHB}
  }
    \subfloat[]{%
    \includegraphics[width=0.32\textwidth]{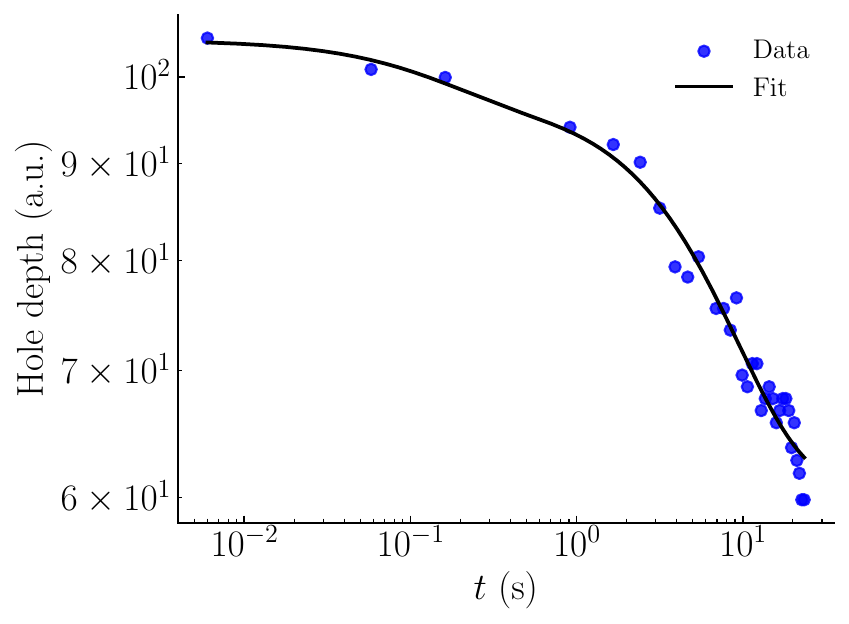}%
    \label{fig:EU-spin lifetime}
  }

  \caption{
    \justifying
    (a) Photoluminescence spectra of Eu$^{3+}$ ions in the Eu$^{3+}$ molecular complex with 100\% concentration measured at room temperature, obtained by exciting the ions into the $^5D_1$ manifold near 532~nm, followed by relaxation to the emitting $^5D_0$ state. The $^5D_0 \rightarrow {}^7F_0$ emission line appears as a weak peak in the spectrum. 
    (b) Photoluminescence excitation of Eu$^{3+}$ ions in the Eu$^{3+}$ complex, illustrating the inhomogeneous broadening of the optical transition. The spectrum is fitted with a Lorentzian function, yielding a full width at half maximum (FWHM) of $28.26 \pm 0.029$ ~GHz. 
    (c) Fluorescence intensity decay of Eu$^{3+}$ ions in the Eu$^{3+}$ complex, revealing the excited-state $^5D_{0}$ lifetime with a characteristic decay constant of $732.85\pm0.15$ $\mu s$. 
    (d) Spectral hole burned in the Eu$^{3+}$ complex, fitted with a Lorentzian function with a FWHM of $10.76\pm0.18$ MHz. 
    (e) Spectral-hole lifetime of Eu$^{3+}$ ions in the Eu$^{3+}$ complex, exhibiting a bi-exponential decay with one short and one long lifetime of $154\pm140$ ms and $8.084\pm1.100$ s, respectively, with corresponding weights of $0.16\pm0.06$ and $0.84\pm0.06$.  }
  \label{fig:2}
\end{figure*}

We first consider the Er$^{3+}$ complex. Figure~\ref{fig:OD-Er-CT} shows the absorption profile of the Er$^{3+}$ sample with 100\% concentration measured at room temperature and at 0.69 K. Here, 100\% concentration means that each molecule in the sample contains one Er$^{3+}$ ion. At room temperature, the spectrum forms a broad feature extending over tens of nanometers, reflecting strong phonon coupling and the associated thermal broadening. Upon cooling the sample to 688~mK, seven principal absorption peaks become resolved. These peaks correspond to transitions from the lowest crystal-field level of the $^4I_{15/2}$ ground-state manifold to the seven crystal-field levels of the $^4I_{13/2}$ excited-state manifold. The largest optical depth is observed near 1530~nm, corresponding to the transition between the lowest crystal-field sublevels of the $^4I_{15/2}$ and $^4I_{13/2}$ manifolds. The inhomogeneous full width at half maximum of this peak is $40.2 \pm 1.1$~GHz. An additional absorption feature is visible near 1536.7~nm, below the energy of the lowest transition originating from the lowest ground-state crystal-field level. Its position is consistent with absorption from the second-lowest ground-state level, $|2\rangle_g$, to the lowest excited-state level, $|1\rangle_e$, as discussed further in Section~\ref{Rethermalization}.

To further resolve the Er$^{3+}$ energy-level structure, we obtained PL spectra, shown in Figure~\ref{fig:PLE}. In this experiment, the Er$^{3+}$ ions were excited from the $^4I_{15/2}$ ground-state manifold to the $^4I_{11/2}$ manifold using light near 980~nm, and the spectra were recorded at 1.8 and 35~K. When combined with the low-temperature absorption data in Figure~\ref{fig:OD-Er-CT}, these PLE measurements enable reconstruction of the crystal-field level scheme of the Er$^{3+}$ complex, as summarized in Figure~\ref{fig:Energy-levels-Er}. From this combined analysis, we identify eight crystal-field levels in the $^4I_{15/2}$ ground-state manifold and seven crystal-field levels in the $^4I_{13/2}$ excited-state manifold. Figure~\ref{fig:Er-lifetime} shows the Er$^{3+}$ fluorescence decay measured at 500~mK. A single-exponential fit yields the $^4I_{13/2}$ excited-state lifetime of $T_1=0.884 \pm 0.013~\mu$s mentioned above.

We next turn to the Tm$^{3+}$ complex. Figure~\ref{fig:OD-Tm-RT} shows the absorption profile of the $^3H_6 \rightarrow {}^3H_4$ transition in the 100\%-concentration Tm$^{3+}$ sample measured at room temperature and at 3.9~K. As in the Er$^{3+}$ case, this room-temperature absorption profile is broadened by phonon coupling. Upon cooling the sample to 3.9~K, the absorption features become significantly narrower and two distinct peaks emerge. These peaks are attributed to transitions from the lowest crystal-field level of the $^3H_6$ ground-state manifold to two crystal-field levels of the $^3H_4$ excited-state manifold. The inhomogeneous full width at half maximum of the lowest-energy absorption peak is $64.4 \pm 2.0$~GHz.

\subsubsection{Eu$^{3+}$}
We finally consider the Eu$^{3+}$ complex. Figure~\ref{fig:Eu-spectrum} shows the PL spectra measured for this sample at room temperature using cw excitation near 532~nm into the $^5D_1$ manifold, followed by relaxation to the emitting $^5D_0$ state. The $^5D_0 \rightarrow {}^7F_0$ emission line appears as a weak feature in the spectrum. From these data, we extract a branching ratio of approximately 1.37\% for the $^5D_{0} \rightarrow {}^7F_{0}$ transition. At 4.2~K, the PLE shown in Figure~\ref{fig:Eu-PL} exhibits an inhomogeneously broadened feature with a linewidth of $28.26 \pm 0.029$ ~GHz centered at around $580.299$ ~nm. We also measured the $^5D_0$ excited-state lifetime at 4.2~K, as shown in Figure~\ref{fig:Eu-T1}, and obtained $T_1 = 732.85 \pm 0.15~\mu$s. This value is comparable to optical lifetimes reported for other Eu$^{3+}$-based materials \cite{kunkel2016dephasing, konz2003temperature, liu2020defect, ahlefeldt2013optical}.

Together, these measurements establish the optical transition energies and crystal-field structure of the investigated complexes and provide the spectroscopic basis for the following sections. We then examine the spectral hole-burning (SHB) response of these complexes.

\subsection{SHB}\label{SHB}

The spectral-hole-burning measurements provide an additional comparison among the three complexes. Under the conditions explored here, persistent spectral holes were observed in the Eu$^{3+}$ complex, whereas no measurable hole was detected in the Er$^{3+}$ and Tm$^{3+}$ complexes. This behavior is consistent with their fundamental differences: Eu$^{3+}$ is a non-Kramers ion with a $J=0$ ground state and nuclear spin in both naturally abundant isotopes, favoring weak magnetic sensitivity and nuclear-spin shelving; Er$^{3+}$ is a Kramers ion with a strong electronic magnetic moment, and only approximately 23\% of natural-abundance Er ions have nonzero nuclear spin; and Tm$^{3+}$, although non-Kramers, has a $J=6$ ground-state manifold and lacks the $J=0$ protection of Eu$^{3+}$. Moreover, the measurements were performed in different thermal environments: the Eu$^{3+}$ experiments used a liquid-helium bath, whereas the Er$^{3+}$ and Tm$^{3+}$ measurements used a cold-plate cryostat. For the Er$^{3+}$ and Tm$^{3+}$ complexes, we attempted to burn a spectral hole using weak optical pulses with an individual pulse duration of 1~ms, while varying the total burn time from a few hundreds of microseconds to several hundred milliseconds. The spectral region around the burn frequency was then scanned over several hundred megahertz to a few gigahertz using the phase modulator. We also systematically varied the burn power from tens of nanowatts to several milliwatts to determine whether a spectral hole could be observed while ensuring that any possible signal was not obscured by power broadening. For the Eu$^{3+}$ complex, the SHB sequence was implemented differently: a 20~ms burn pulse was followed by a 5~ms waiting time, after which the spectrum was probed over a 200~MHz range during a 2~ms read pulse generated using the AOM.

\subsubsection{Er$^{3+}$ and Tm$^{3+}$}

We performed SHB measurements on Er$^{3+}$- and Tm$^{3+}$-based complexes with concentrations ranging from 1\% to 100\% under external magnetic fields up to 3~T. In all cases, no measurable spectral hole was observed. The energy-level structures of the Er\(^{3+}\) and Tm\(^{3+}\) complexes provide efficient channels for vibrationally assisted nonradiative relaxation. For Er\(^{3+}\), the \({}^4I_{13/2}\rightarrow{}^4I_{15/2}\) energy separation near \(6500~\mathrm{cm^{-1}}\) lies close to the first-overtone region of the ligand C–H stretching modes (\(5800-6200~\mathrm{cm^{-1}}\)). This assignment is consistent with the short optical lifetime of \(T_{1,\mathrm{opt}}\approx0.88~\mu\mathrm{s}\) measured here. For Tm\(^{3+}\), the \({}^3H_4\rightarrow{}^3H_5\) separation of approximately \(4348~\mathrm{cm^{-1}}\) overlaps the C–H stretch + C–H bend combination-band region (\(4200-4500~\mathrm{cm^{-1}}\)), suggesting that ligand-assisted multiphonon relaxation may also be efficient in this system.

Efficient nonradiative relaxation alone, however, does not preclude spectral-hole burning. Persistent SHB requires both efficient transfer of the optically selected population into a sufficiently long-lived shelving state and preservation of the spectral selectivity created by the burn pulse. The short Er\(^{3+}\) excited-state lifetime demonstrates rapid decay of the optically excited manifold, but does not determine the branching ratio into possible shelving states. If the dominant nonradiative pathways preferentially return population to the initially absorbing state, or if the shelving population relaxes rapidly, the resulting hole would have low contrast. The measured Er\(^{3+}\) lifetime corresponds to a lifetime-limited linewidth contribution of only \(\sim0.18\) MHz, so lifetime broadening alone cannot account for the absence of a resolvable feature over the much larger frequency ranges explored here.

The molecular vibrational environment may additionally affect both coherence and shelving dynamics. Molecular vibrations modify the ligand geometry and therefore modulate the crystal-field parameters and the effective spin Hamiltonian. Consequently, even high-energy intramolecular modes can possess substantial spin–phonon coupling coefficients, although their contribution to equilibrium spin relaxation at cryogenic temperature is strongly restricted by their negligible thermal occupation and large energy mismatch with the low-energy spin and hyperfine transitions. Under optical burning, however, ligand-assisted nonradiative relaxation creates a nonequilibrium population of molecular vibrations. These excited high-energy modes can directly modulate crystal-field and spin-Hamiltonian parameters while they are populated and may contribute through higher-order or occupation-dependent spin–phonon processes \cite{garlatti2020unveiling}. In addition, anharmonic vibrational relaxation can redistribute their energy into lower-frequency vibrational modes; such high-to-low-energy vibrational cooling has been studied explicitly in low-temperature molecular crystals and following selective C–H stretch excitation in molecular solids~\cite{hill1988vibrational,hill1988model,lu2019abinitio}. Low-energy molecular modes are particularly effective at driving spin relaxation and dephasing because they are thermally accessible and overlap the energy scales relevant to spin and crystal-field dynamics~\cite{garlatti2023critical}. Thus, strong optical multiphonon relaxation may affect SHB through both direct nonequilibrium modulation of the local spin/crystal-field environment and indirect population of lower-energy vibrational modes. The magnitude of either contribution under the present burn conditions is not determined by the present measurements.

\begin{figure*}[!t]
  \centering

  \subfloat[]{%
    \includegraphics[width=0.32\textwidth]{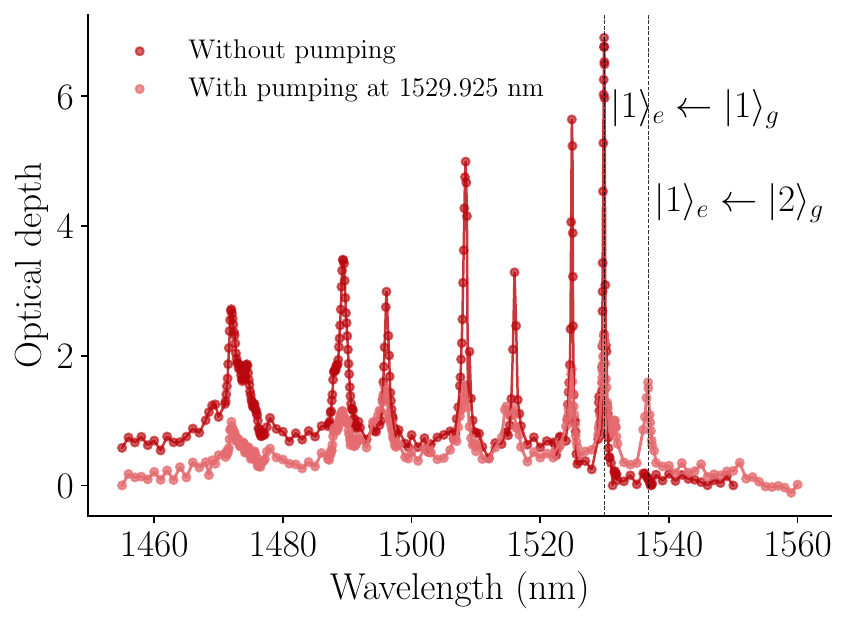}%
    \label{fig:OD-Er-pump1530}}
   \subfloat[]{%
    \includegraphics[width=0.32\textwidth]{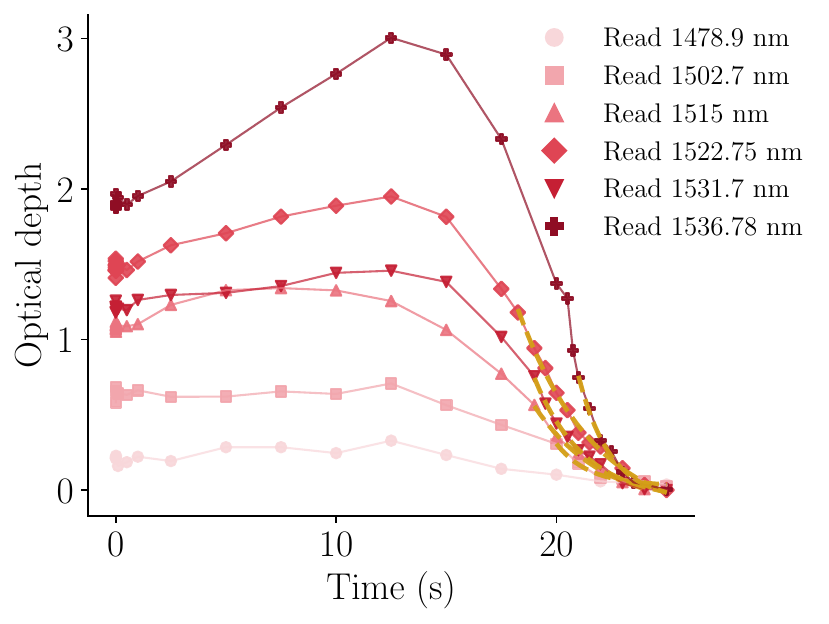}%
    \label{fig:OD-Er-lifetime-pump1530}
  }
     \subfloat[]{%
    \includegraphics[width=0.32\textwidth]{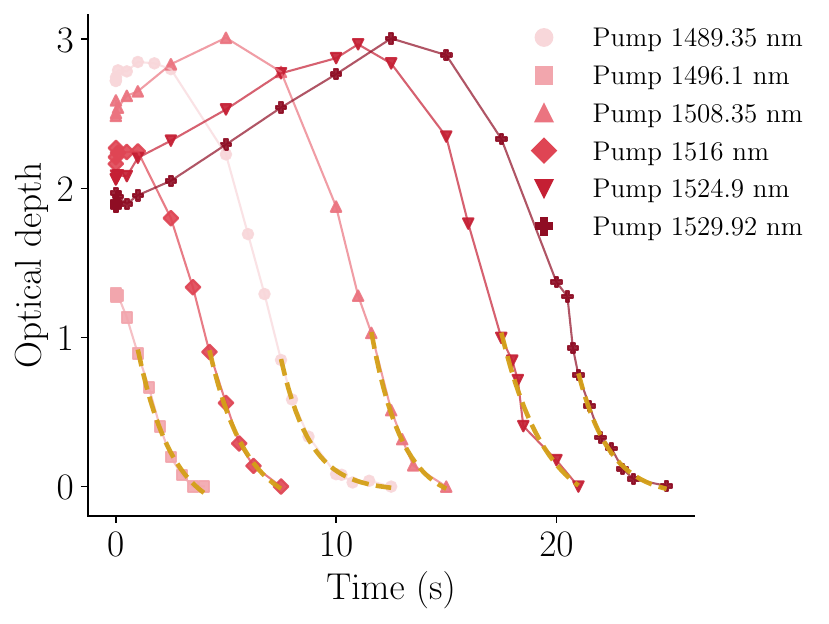}%
    \label{fig:OD-Er-lifetime-pumpall}
  }

  \subfloat[]{%
    \includegraphics[width=0.32\textwidth]{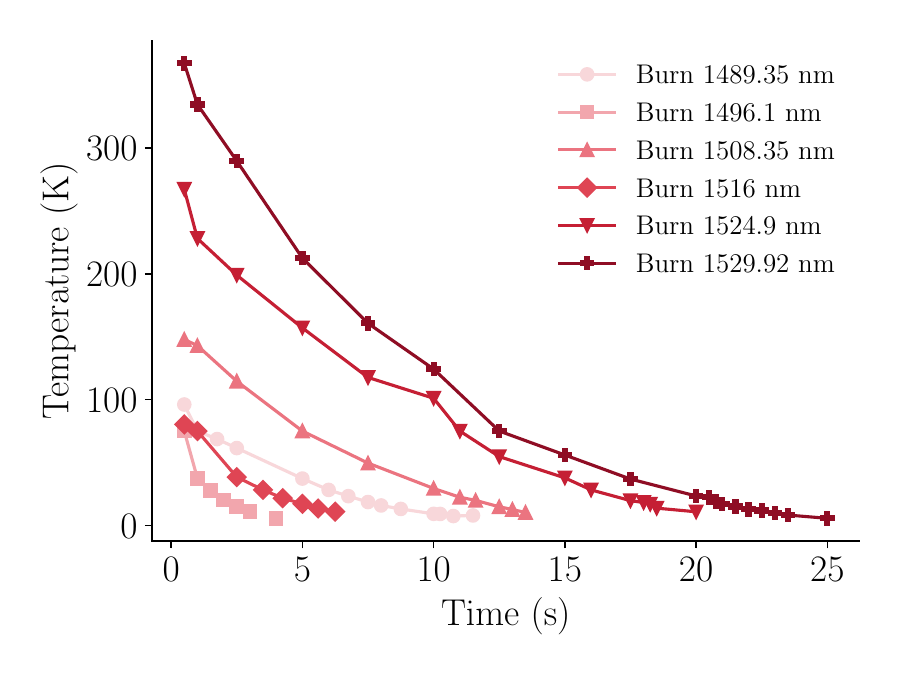}%
    \label{fig:Temp_vs_time}
   }  \subfloat[]{%
  \includegraphics[width=0.3\textwidth]{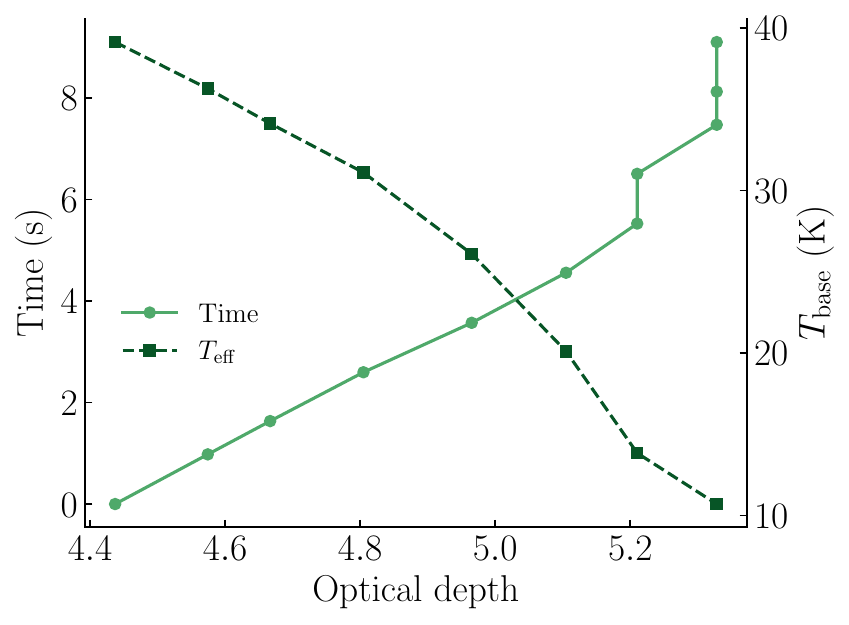}%
    \label{fig:Tm-Lifetime-0T-100percent-3.9K}
  }
   \subfloat[]{%
    \includegraphics[width=0.32\textwidth]{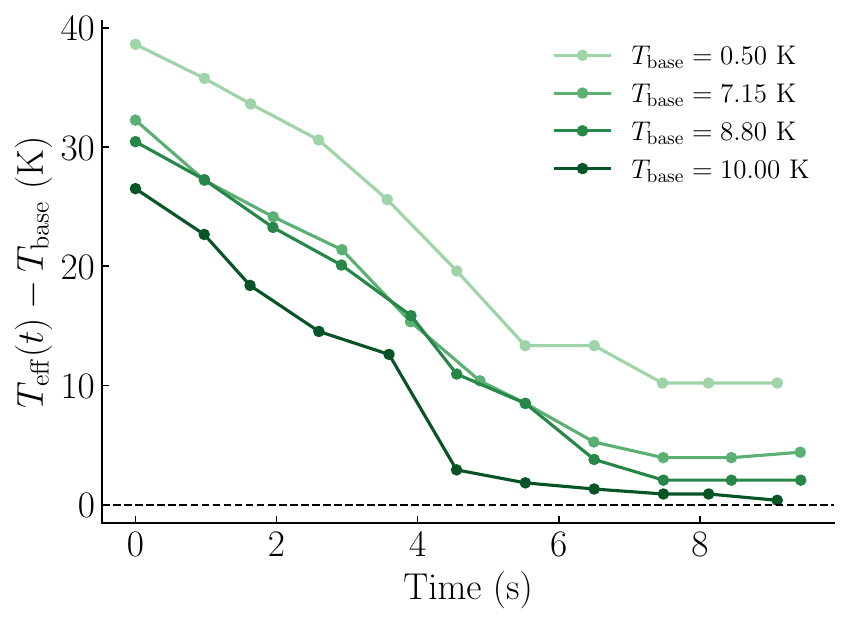}%
    \label{fig:Tm-temperature-vs-time}
  }
  

  \caption{
    \justifying
    (a) The dark red curve shows the absorption profile of the Er$^{3+}$ complex without pumping, while the light red curve shows the absorption profile measured 50 ms after pumping the population from the lowest level of $^4I_{15/2}$ ($|1\rangle_g$) using a laser pulse at 1529.92 nm. 
    (b) The population is pumped from ${|1\rangle_g}$ via the ${|1\rangle_g} \rightarrow {|1\rangle_e}$ transition (1529.925 nm), and the population in ${|2\rangle_g}$ is subsequently probed by driving different transitions, indicated by different colors. The tail (last few seconds) of each trace is fitted with a single exponential, and the extracted values are reported in Section~\ref{Rethermalization}. 
    (c) The population is pumped from ${|1\rangle_g}$ to ${|i\rangle_e}$ (i = 1, ..., 7), as indicated by different colors. The population in ${|2\rangle_g}$ is then probed via the ${|2\rangle_g} \rightarrow {|1\rangle_e}$ transition. The tail of each trace is fitted with a single exponential, and the corresponding decay times are listed in Section~\ref{Rethermalization}. 
    (d) The temperature evolution after pumping, inferred from the Boltzmann population distribution and the optical-depth transients in Figure~\ref{fig:OD-Er-lifetime-pumpall}. The largest inferred temperatures are upper bounds because temperature-dependent broadening of the probed transition is not included in this conversion.
    (e) Relationship between the measured optical depth and the waiting time after a 100~ms, $\sim 15$~mW pump pulse, OD$(t)$ (light-green circles and solid line, left axis), together with the equilibrium optical-depth calibration as a function of temperature, OD$(T)$ (dark-green squares and dashed line, right axis), for the Tm$^{3+}$ complex. The comparison of these two data sets provides a direct mapping from OD$(t)$ to the effective local temperature.
    (f) Temperature evolution of the optically excited spatial volume in the Tm$^{3+}$ complex after optical pumping, extracted at different cryostat base temperatures by mapping the transient optical depth onto the measured equilibrium OD$(T)$ calibration.
  }
  \label{fig:er_tm_rethermalization}
\end{figure*}
Additional factors likely contribute to the absence of SHB signals. All our samples, including those used for Er$^{3+}$ and Tm$^{3+}$ measurements were performed on powdered samples rather than single crystals. In a powder sample, heat transport is further hindered by the weak thermal contact between neighboring crystallites, which touch only at small contact points, as well as by the presence of vacuum gaps between particles. As a result, optically-induced heat is removed less efficiently than in a bulk crystal, leading to stronger local heating and increased phonon populations. Because molecular crystals also generally exhibit lower thermal conductivity than typical inorganic hosts, these phonons can remain trapped for longer times and thereby enhance nonradiative relaxation. The use of a cold-plate cryostat for the Er$^{3+}$ and Tm$^{3+}$ measurements---with substantially lower cooling power than a liquid-helium bath---further strengthens this thermal bottleneck, which can broaden both the spectral hole and the homogeneous linewidth and thereby make the hole more difficult to detect. This consideration also motivated the local-heating measurements discussed in Subsection~\ref{Rethermalization}. 

Finally, magnetic interactions within the molecular solid may also play an important role. In the concentrated Er$^{3+}$ samples, magnetic dipole--dipole coupling between neighboring Er$^{3+}$ ions can promote spin flip-flops and spectral diffusion, thereby broadening the optical transition and shortening the population-storage lifetime. In addition, two coordinating nitrogen nuclei in each complex possess nonzero nuclear spin. For Kramers ions such as Er$^{3+}$, coupling to these nearby nuclear spins may further broaden the spectral hole and homogeneous linewidth while shortening the spin lifetime.

\subsubsection{Eu$^{3+}$}

In contrast, the Eu$^{3+}$ complex behaves differently for several reasons. Our SHB measurements target the ${}^{5}D_{0} \leftarrow {}^7F_{0}$ transition near $580~\mathrm{nm}$ ($\sim 17{,}200~\mathrm{cm^{-1}}$). Even lower-lying relevant energy separations, such as the ${}^{5}D_{0} \leftarrow {}^7F_{6}$ gap, remain on the order of $820~\mathrm{nm}$ ($\sim 12{,}195~\mathrm{cm^{-1}}$). These energy gaps are much larger than the C--H overtone and combination-band energies discussed above, so multiphonon relaxation is expected to be less efficient. The measured ${}^{5}D_0$ lifetime of $732.85 \pm 0.15~\mu$s is comparable to values reported for other Eu$^{3+}$-based materials~\cite{kunkel2016dephasing,konz2003temperature,liu2020defect,ahlefeldt2013optical}. However, this comparison alone does not determine the relative contributions of radiative and nonradiative decay. For Eu$^{3+}$, the radiative lifetime can be estimated from the integrated, spectrally corrected emission bands using the ${}^{5}D_0\rightarrow{}^{7}F_1$ transition as a purely magnetic-dipole internal reference and an appropriate refractive index. Comparing this radiative lifetime with the measured lifetime then allows the intrinsic Eu$^{3+}$ emission quantum efficiency to be estimated without an independent quantum-yield measurement. Such an analysis is not performed here, so we do not quantify the nonradiative contribution. In addition, the Eu$^{3+}$ measurements were performed in a liquid-helium bath, which provides significantly higher cooling power and therefore better suppresses phonon-induced decoherence. Finally, Eu$^{3+}$ is a non-Kramers ion with zero electronic spin in the ground state ($J=0$), leading to much weaker magnetic interactions with nearby nuclear spins. Together, these factors enable the observation of spectral hole burning in the Eu$^{3+}$ complex, in contrast to the Er$^{3+}$ and Tm$^{3+}$ complexes.

Figure~\ref{fig:Eu-SHB} shows a representative SHB signal obtained in the Eu$^{3+}$ complex. The measured full width at half maximum (FWHM) of the spectral hole is $\gamma_{\mathrm{hole}} = 11.40 \pm 0.04$~MHz, which sets an upper bound on the homogeneous linewidth of $\Gamma_h \leq \gamma_{\mathrm{hole}}/2 \approx 5.7$~MHz. In addition, we measured the nuclear spin lifetime of the Eu$^{3+}$ ions. As shown in Figure~\ref{fig:EU-spin lifetime}, the population decay is well described by two characteristic time scales: a shorter component of $154 \pm 140$~ms and a longer component of $8.084 \pm 1.100$~s, with corresponding amplitudes of $0.16 \pm 0.06$ and $0.84 \pm 0.06$, respectively. These results indicate that approximately 16\% of the ions relax on a time scale of about 154~ms, whereas the majority (around 84\%) relax on a time scale of about 8~s. Alternatively, the biexponential behavior may arise because different nuclear-spin states relax at different rates.

\subsection{Laser-Induced Local Heating Dynamics}\label{Rethermalization}

Next we examine whether reduced thermalization and laser-induced local heating could help explain why SHB is not observed in Er$^{3+}$ and Tm$^{3+}$. We first characterize this behavior in the Er$^{3+}$ complex by comparing the absorption spectrum before and after optical pumping. The absorption profile was recorded using a weak 100~$\mu$s probe pulse, yielding the dark red curve in Figure~\ref{fig:OD-Er-pump1530}. In a second measurement, the 100\%-concentration Er$^{3+}$ sample was pumped on the $|1\rangle_e \leftarrow |1\rangle_g$ transition at 1529.925~nm, allowed to evolve for 50~ms, and then probed again with the same weak 100~$\mu$s pulse, yielding the light red curve in Figure~\ref{fig:OD-Er-pump1530}. Relative to the original seven peaks, the additional transitions that appear in the orange curve are shifted by approximately 870~GHz to lower frequency. This offset is consistent with transitions originating from the second-lowest crystal-field level of the ground-state manifold, $|2\rangle_g$ of $^4I_{15/2}$, to the crystal-field levels $|i\rangle_e$ ($i=1,\ldots,7$) of the excited-state manifold $^4I_{13/2}$. We therefore attribute the appearance of these extra peaks to slow rethermalization, which leaves a non-negligible population in $|2\rangle_g$ after optical pumping. 

To further investigate the dynamics of the $|2\rangle_g$ level, we performed two additional measurements. In the first experiment, Er$^{3+}$ ions in the 100\%-concentration sample were excited from $|1\rangle_g$ to the lowest crystal-field level of the excited-state manifold, $|1\rangle_e$, using a 1.3~s laser pulse at 1529.925~nm with a peak power of $\sim 12$~mW. After a waiting time of 50~ms, the population in $|2\rangle_g$ was probed using a 100~$\mu$s read pulse with a peak power of $\sim 350$~$\mu$W at wavelengths corresponding to the $|i\rangle_e \leftarrow |2\rangle_g$ transitions ($i=1,\ldots,6$) within the $^4I_{13/2}$ manifold. As shown in Figure~\ref{fig:OD-Er-lifetime-pump1530}, the population in $|2\rangle_g$ decays as the sample thermalizes. The tail of each trace was fitted with a single-exponential decay, except for the $|6\rangle_e \leftarrow |2\rangle_g$ transition, which did not provide sufficient contrast for reliable fitting. The extracted decay times and relative transition strengths inferred from the measured optical depths are summarized in Table~\ref{tab:er-rethermalization-transitions}.

\begin{table}[t]
\centering
\small
\resizebox{\columnwidth}{!}{%
\begin{tabular}{cccc}
\toprule
Transition & Wavelength (nm) & $\tau$ (s) & Relative strength \\
\midrule
$|1\rangle_e \leftarrow |2\rangle_g$ & 1536.78 & $1.4 \pm 0.2$ & 1.00 \\
$|2\rangle_e \leftarrow |2\rangle_g$ & 1531.70 & $2.3 \pm 0.2$ & 0.49 \\
$|3\rangle_e \leftarrow |2\rangle_g$ & 1522.75 & $3.5 \pm 0.3$ & 0.65 \\
$|4\rangle_e \leftarrow |2\rangle_g$ & 1515.00 & $3.1 \pm 0.5$ & 0.42 \\
$|5\rangle_e \leftarrow |2\rangle_g$ & 1502.70 & $1.4 \pm 0.3$ & 0.24 \\
$|6\rangle_e \leftarrow |2\rangle_g$ & 1478.90 & --- & 0.11 \\
\bottomrule
\end{tabular}
}
\caption{Decay times of the $|2\rangle_g$ population measured through different excited-state transitions and the corresponding relative transition strengths. No reliable decay time could be extracted for the $|6\rangle_e \leftarrow |2\rangle_g$ transition because of its low contrast.}
\label{tab:er-rethermalization-transitions}
\end{table}

In the second experiment, we investigated the population dynamics of $|2\rangle_g$ for different excitation wavelengths. Er$^{3+}$ ions were excited from $|1\rangle_g$ to $|i\rangle_e$ ($i=1,\ldots,6$) at the corresponding transition wavelengths, with the duration adjusted for each transition (Table~\ref{chap5-table1}). After a waiting time of 50~ms, the population in $|2\rangle_g$ was probed using a 100~$\mu$s pulse at 1536.78~nm, corresponding to the $|1\rangle_e \leftarrow |2\rangle_g$ transition, with a peak power of $\sim 350$~$\mu$W. For each excitation wavelength, the pump duration was adjusted to produce a clearly measurable readout signal.


\begin{table}[t]
\centering
\caption{Transitions and Pump Pulse Durations - T = 3.85 K}
\begin{tabular}{cc}
\toprule
Transitions & Pump Pulse Duration \\
\midrule
1529.92 nm & 1.300 s \\
1524.9 nm & 930 ms \\
1516 nm & 8 ms \\
1508.35 nm & 250 ms \\
1496.1 nm & 300 us \\
1489.35 nm & 166 ms \\
\bottomrule
\end{tabular}
\label{chap5-table1}
\end{table}

Figure~\ref{fig:OD-Er-lifetime-pumpall} shows the population dynamics of the $|2\rangle_g$ level following optical pumping, from which the local rethermalization dynamics can be inferred. Immediately after the pump pulse, the locally illuminated region of the sample is heated, which increases the thermal population of higher crystal-field levels in the $^4I_{15/2}$ ground-state manifold, $|i\rangle_g$ with $i=2,3,\ldots,8$. Assuming quasi-thermal equilibrium within the manifold, the population of each level is described by the Boltzmann distribution,
\begin{equation}
N_{|i\rangle_g}(T) = \frac{g_i e^{-E_i/(k_B T)}}{\sum_{j=1}^{N} g_j e^{-E_j/(k_B T)}},
\end{equation}\label{eq:boltzmandist}
where $N_{|i\rangle_g}(T)$ is the population of the level $|i\rangle_g$ at temperature $T$, $E_i$ is its energy, $g_i$ is its degeneracy, and $k_B$ is the Boltzmann constant. Because for the Er$^{3+}$ complex the crystal-field energies are known from Figure~\ref{fig:Energy-levels-Er}, this expression allows us to calculate the population of $|2\rangle_g$ as a function of temperature. Independently, Figure~\ref{fig:OD-Er-lifetime-pumpall} provides the time evolution of the optical depth associated with the $|1\rangle_e \leftarrow |2\rangle_g$ transition at 1536.78~nm after optical pumping, which directly tracks the population dynamics of $|2\rangle_g$.

\begin{figure*}[htb]
  \centering

  \subfloat[]{%
    \includegraphics[width=0.32\textwidth]{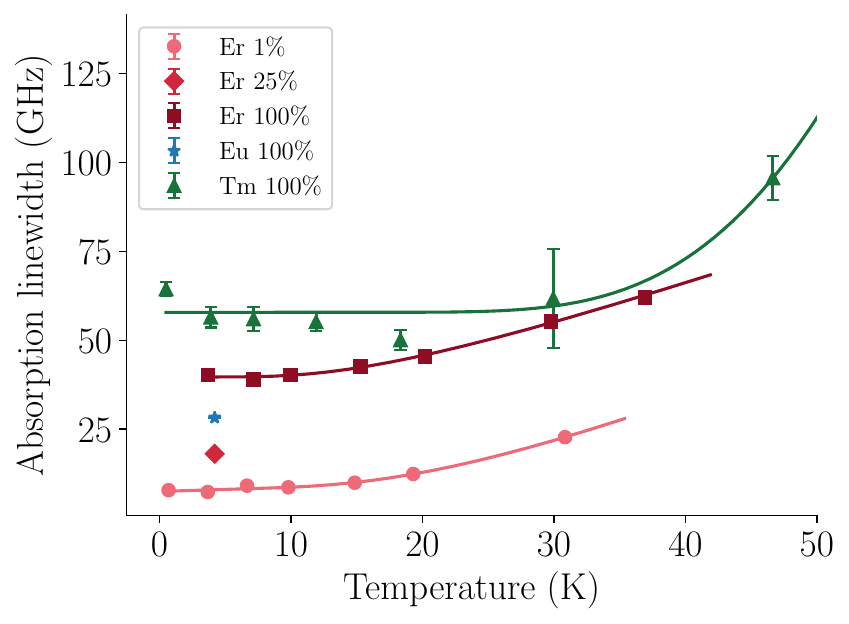}%
    \label{fig:Inhom-vs-temp-0T}}
  \subfloat[]{%
    \includegraphics[width=0.32\textwidth]{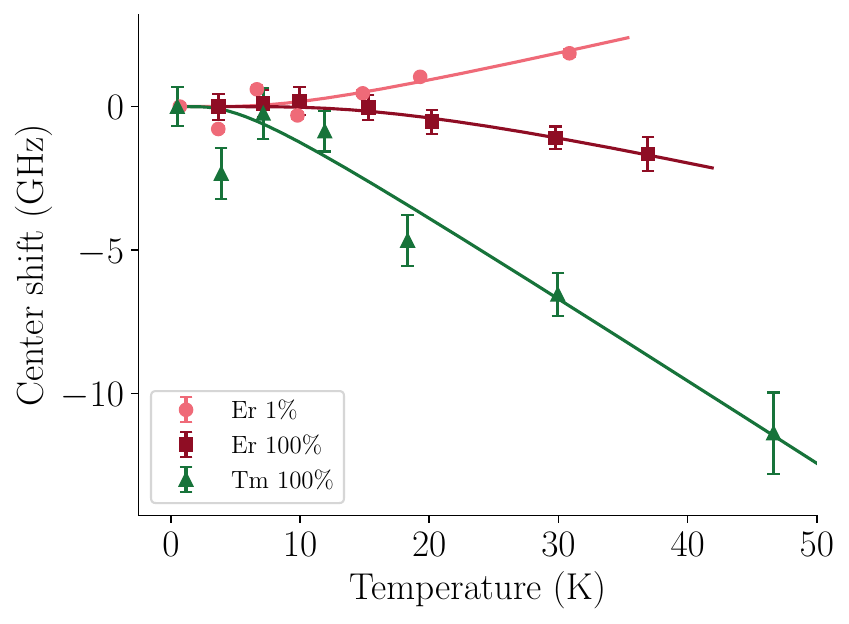}%
    \label{fig:center-vs-temp-0T}}
  
  \caption{
    \justifying
    (a) Temperature dependency of the absorption linewidth for Er$^{3+}$-acac complex with 100\% concentration (dark red), Er$^{3+}$-acac sample with 1\% concentration (light red), and Tm$^{3+}$-acac complex with 100\% concentration (green). Data points are fitted with this model $\Gamma(T) = \Gamma_{inhom} + \alpha_{\mathrm{TLS}} T + \alpha_{\mathrm{w}} \frac{1}{e^{\Delta E_{\mathrm{}}/(k_B T)} - 1 }$\cite{bottger2016optical}
    . (b) Thermal shift of transition energy for Er$^{3+}$-acac complex with 100\% concentration (dark red), Er$^{3+}$-acac sample with 1\% concentration (light red), and Tm$^{3+}$-acac complex with 100\% concentration (green). Data points are fitted with this model $E(T) = E_0 + \alpha_{\mathrm{s}} \frac{1}{e^{\Delta E_{\mathrm{}}/(k_B T)} - 1 }$\cite{bottger2016optical}.
  }
  \label{fig4}
\end{figure*}

\begin{table*}[t]
\centering
\renewcommand{\arraystretch}{1.4}
\resizebox{\textwidth}{!}{%
\begin{tabular}{|c|c|c|c|c|c|c|}
\hline
 & $\Gamma_{inhom}^*$ & $\alpha_{\mathrm{TLS}}^*$ & $\alpha_{\mathrm{w}}^*$ & $\Delta E^*$ & $\alpha_{\mathrm{s}}^{**}$ & $\Delta E^{**}$ \\ \hline
Er$^{3+}$ 1\% & \shortstack{$7.513 \pm 0.224$~GHz \\ $0.2506 \pm 0.0075$~cm$^{-1}$} & \shortstack{$0.110 \pm 0.038$~GHz/K \\ $0.00367 \pm 0.00127$~cm$^{-1}$/K} & \shortstack{$107.212 \pm 29.091$~GHz \\ $3.58 \pm 0.97$~cm$^{-1}$} & \shortstack{$1480.718 \pm 191.255$~GHz \\ $49.39 \pm 6.38$~cm$^{-1}$} & \shortstack{$3.084 \pm 1.151$~GHz \\ $0.103 \pm 0.038$~cm$^{-1}$} & \shortstack{$609.423 \pm 131.185$~GHz \\ $20.33 \pm 4.38$~cm$^{-1}$} \\ \hline
Er$^{3+}$ 100\% & \shortstack{$39.696 \pm 2.271$~GHz \\ $1.324 \pm 0.076$~cm$^{-1}$} & \shortstack{$0.000 \pm 0.373$~GHz/K \\ $0.000 \pm 0.012$~cm$^{-1}$/K} & \shortstack{$63.688 \pm 36.069$~GHz \\ $2.12 \pm 1.20$~cm$^{-1}$} & \shortstack{$1020.335 \pm 540.403$~GHz \\ $34.04 \pm 18.03$~cm$^{-1}$} & \shortstack{$-5.851 \pm 10.842$~GHz \\ $-0.195 \pm 0.362$~cm$^{-1}$} & \shortstack{$1151.851 \pm 998.677$~GHz \\ $38.42 \pm 33.31$~cm$^{-1}$} \\ \hline
Tm$^{3+}$ 100\% & \shortstack{$57.855 \pm 1.747$~GHz \\ $1.930 \pm 0.058$~cm$^{-1}$} & \shortstack{$0.000 \pm 0.183$~GHz/K \\ $0.000 \pm 0.006$~cm$^{-1}$/K} & \shortstack{$9234.973 \pm 140626.170$~GHz \\ $308 \pm 4691$~cm$^{-1}$} & \shortstack{$5349.533 \pm 14794.701$~GHz \\ $178 \pm 494$~cm$^{-1}$} & \shortstack{$-4.349 \pm 3.061$~GHz \\ $-0.145 \pm 0.102$~cm$^{-1}$} & \shortstack{$312.662 \pm 172.401$~GHz \\ $10.43 \pm 5.75$~cm$^{-1}$} \\ \hline
\end{tabular}%
}
\caption{Fitting parameters of the models in equation~\ref{eq:3} ($^*$) and equation~\ref{eq:4} ($^{**}$). Frequency-valued parameters are reported in both GHz and cm$^{-1}$.}
\label{tab:combined_inhom_center_vs_temp_fit}
\end{table*}

This correspondence between the thermal population of $|2\rangle_g$ predicted by the Boltzmann distribution and the experimentally measured population dynamics of this level in Figure~\ref{fig:OD-Er-lifetime-pumpall} provides a way to convert the optical-depth transient into the local temperature dynamics of the optically excited region, as shown in Figure~\ref{fig:Temp_vs_time}. As the sample cools after the pump pulse, population relaxes from higher-lying crystal-field levels toward lower ones, causing the optical depth of the $|2\rangle_g \rightarrow |1\rangle_e$ transition to first increase and reach a maximum. Upon further cooling, the population gradually returns to the lowest ground-state level, $|1\rangle_g$, and the optical depth decreases toward its equilibrium value. In this way, the transient response of the $|2\rangle_g \rightarrow |1\rangle_e$ absorption line serves as a local thermometer for the laser-heated region of the sample.  

This conversion for Er$^{3+}$ assumes that the measured peak optical depth changes only because of the Boltzmann redistribution of population among the crystal-field levels. In practice, increasing temperature also broadens the optical transition and thereby reduces its peak optical depth. Because this broadening contribution is not removed in the present analysis, the full reduction in optical depth is attributed to population redistribution, leading to an overestimate of the temperature. The largest values in Figure~\ref{fig:Temp_vs_time}, including the one above 300~K, should therefore be interpreted as upper bounds; accounting for the linewidth broadening would yield lower temperatures.

We also investigated rethermalization in the 100\%-concentration Tm$^{3+}$ complex. In this experiment, the population was pumped out of the $^3H_6$ ground-state manifold for 100~ms using an optical power of $\sim 15$~mW, followed by a variable waiting time. The system was then probed with a weak 100~$\mu$s pulse at the same wavelength, and the transmitted intensity was recorded on the oscilloscope. Repeating this sequence for different waiting times gives the transient optical depth OD$(t)$, shown by the light-green circles and solid line in Figure~\ref{fig:Tm-Lifetime-0T-100percent-3.9K}. The optical depth increases with waiting time as the laser-heated region cools and rethermalizes; a single-exponential fit gives a characteristic time constant of $\tau = 2.39 \pm 0.21$~s. Independently, we measured the equilibrium optical depth at different cryostat temperatures to obtain OD$(T)$, shown by the dark-green squares and dashed line in the same panel.

For the Tm$^{3+}$ complex, the temperature conversion is therefore more direct: the measured OD$(T)$ curve in Figure~\ref{fig:Tm-Lifetime-0T-100percent-3.9K} serves as an empirical calibration for converting each value of the transient OD$(t)$ into an effective local temperature. Applying this mapping gives the temperature evolution shown in Figure~\ref{fig:Tm-temperature-vs-time}. Because the measured OD$(T)$ calibration already contains both the Boltzmann redistribution of population and the reduction in peak optical depth caused by temperature-dependent linewidth broadening, these two contributions do not need to be separated, making the extracted Tm$^{3+}$ temperature dynamics more reliable than the Boltzmann-only estimates for Er$^{3+}$. However, at low temperatures the OD$(T)$ calibration becomes nearly flat, reducing the sensitivity of the optical thermometer. In this regime, small uncertainties in optical depth produce large uncertainties in the inferred temperature. Therefore, once the late-time OD$(t)$ agrees with the equilibrium optical depth at the base temperature within the experimental resolution, the sample should be regarded as having returned to the base temperature within the resolution of the thermometer, rather than interpreting a small residual inferred temperature rise quantitatively. As a complementary estimate, assuming that the two lowest crystal-field levels of the $^3H_6$ ground-state manifold dominate over the measured temperature range, fitting OD$(T)$ using Eq.~\ref{eq:boltzmandist} with $j\in\{1,2\}$ gives $\Delta E_{21}=E_2-E_1=44.1 \pm 0.7$~cm$^{-1}$. Repeating the transient measurement at different cryostat base temperatures shows that the laser-induced temperature rise decreases as the base temperature increases. For comparable pump conditions, this smaller temperature excursion indicates a larger effective heat capacity at higher temperature.

These results show that optical excitation can drive significant local heating and slow thermal relaxation in these powder-like RE molecular complexes, especially under high-power or long-duration pumping conditions. They also allow us to estimate the level of heating based on the pump-pulse parameters. As a result, the effective heating maps in the Supporting Information (panels b and c) indicate that the optical powers and pulse durations used for our SHB burn measurements lie in a low-heating regime, where the local temperature is expected to remain close to the cryostat base temperature. Therefore, while laser-induced heating is an important constraint in these powder samples, the absence of measurable SHB in the Er$^{3+}$ and Tm$^{3+}$ complexes is unlikely to be explained solely by insufficient cooling during the burn sequence. Motivated by this temperature sensitivity, we next examine how the transition energy and inhomogeneous linewidth evolve with temperature.


\subsection{Temperature Dependence of the optical absorption Linewidth and Energy}\label{Linewidth Broadening vs Temperature}

Next we further investigate how temperature, whether induced by heating or set by the cryostat, affects the spectroscopic properties of the REI complexes. To begin, we measured the absorption linewidth as a function of cryostat temperature for the Er$^{3+}$ complexes at 1\% and 100\% concentration and for the 100\%-concentration Tm$^{3+}$ complex. The resulting temperature dependence is shown in Figure~\ref{fig:Inhom-vs-temp-0T}. The data was fitted using the model \cite{bottger2016optical}

\begin{equation}
    \Gamma(T) = \Gamma_{inhom} + \alpha_{\mathrm{TLS}} T + \alpha_{\mathrm{w}} \frac{1}{e^{\Delta E_{\mathrm{}}/(k_B T)} - 1 },
    \label{eq:3}
\end{equation}
where $\Gamma_{inhom}$ is the temperature-independent inhomogeneous contribution to the linewidth, $\alpha_{\mathrm{TLS}}$ is the coefficient associated with the linear temperature dependence, $\alpha_{\mathrm{w}}$ is the phonon coupling constant, and $\Delta E$ is the energy separation relevant to the resonant phonon-mode contribution, which in the present case can be associated with crystal-field splittings in the ground- and excited-state manifolds.

The fitting parameters obtained from Eq.~\ref{eq:3} are summarized in Table~\ref{tab:combined_inhom_center_vs_temp_fit} for the Er$^{3+}$ complexes at 1\% and 100\% concentration and for the 100\%-concentration Tm$^{3+}$ complex. For the 100\%-concentration Er$^{3+}$ complex, the fitted value of $\Delta E$ is comparable to the $|2\rangle_g \leftrightarrow |1\rangle_g $ ground-state splitting ($\sim 870$~GHz). 

For the 1\%-concentration Er$^{3+}$ complex, the fitted value is $1480.718 \pm 191.255$~GHz. We note that the fitted $\Delta E$ should be regarded as an effective energy scale arising from a broadened distribution of energy splittings rather than as the exact value of a single crystal-field separation \cite{young1966apparent}. For the 100\%-concentration Tm$^{3+}$ complex, the uncertainties in $\alpha_{\mathrm{w}}$ and $\Delta E$ are relatively large, likely because of the limited number of data points and the more rapid variation of the linewidth at higher temperatures.

In addition to the linewidth broadening, we observe a temperature-dependent shift of the transition energy. The measured thermal shift for the Er$^{3+}$ complexes at 1\% and 100\% concentration and for the 100\%-concentration Tm$^{3+}$ complex is shown in Figure~\ref{fig:center-vs-temp-0T}. This data was fitted using the model \cite{bottger2016optical}

\begin{equation}
    E(T) = E_0 + \alpha_{\mathrm{s}} \frac{1}{e^{\Delta E_{\mathrm{}}/(k_B T)} - 1 },
    \label{eq:4}
\end{equation}
where $E_0$ is the transition energy at $T = 0$~K, $\alpha_s$ is a coupling coefficient that quantifies the strength of the electron--phonon interaction \cite{mccumber1963linewidth}, and $\Delta E$ is defined as in Eq.~\ref{eq:3}. The corresponding fitting parameters are listed in Table~\ref{tab:combined_inhom_center_vs_temp_fit}. For the Er$^{3+}$ complexes at 1\% and 100\% concentration, the fitted values of $\Delta E$ are approximately 609 and 1151~GHz, respectively. As in the linewidth analysis, these reduced fitted values are consistent with interpreting $\Delta E$ as an effective energy scale rather than as the exact value of a single crystal-field splitting. For the Tm$^{3+}$ complex, the fitted value is approximately 313~GHz. However, because the full crystal-field level scheme of the Tm$^{3+}$ complex has not been resolved, we cannot determine whether this value corresponds to a specific crystal-field splitting.

\section{Discussion and Conclusion}

In this work, we investigated the low-temperature optical properties of rare-earth molecular complexes of the form $[\mathrm{Ln}(acac)_3(phen)]$, with $\mathrm{Ln}=\mathrm{Er}^{3+}$, $\mathrm{Tm}^{3+}$, and $\mathrm{Eu}^{3+}$. Our measurements establish that these molecular systems remain optically addressable at cryogenic temperature and exhibit different spectroscopic behavior depending on the rare-earth ion. For the Er$^{3+}$ complex, low-temperature absorption and PLE measurements allowed us to resolve the crystal-field structure of the lowest two manifolds and reconstruct the corresponding energy-level scheme. For the Tm$^{3+}$ complex, we resolved two absorption peaks within the $^3H_6 \rightarrow {}^3H_4$ transition at 3.9~K and measured an inhomogeneous FWHM of $64.4 \pm 2.0$~GHz for the lowest-energy peak. For the Eu$^{3+}$ complex, we measured the branching ratio, the inhomogeneous linewidth, and the optical lifetime, and we observed spectral hole burning (SHB) together with a long spin lifetime. These measurements enabled us to place an upper bound on the homogeneous linewidth.

A central result of this study is the clear contrast between the Eu$^{3+}$ complex and the Er$^{3+}$ and Tm$^{3+}$ complexes. SHB was observed in Eu$^{3+}$, whereas no spectral hole was resolved in Er$^{3+}$ or Tm$^{3+}$ under the conditions explored here. We first considered whether this contrast could simply arise from the different cryogenic setups and the associated local heating. However, the effective thermal analysis summarized in panel (b) of the effective thermal parameter maps in the Supporting Information indicates that, for the range of optical powers and burn durations used in our experiments---from tens of nW to mW and from several $\mu$s to seconds---the low-power, short-duration conditions relevant to SHB remain in the low-temperature regime. This suggests that the absence of SHB in Er$^{3+}$ and Tm$^{3+}$ cannot be explained solely by insufficient cooling.

We next consider ion-dependent magnetic and vibrational relaxation processes. Eu$^{3+}$ is a non-Kramers ion with a $J=0$ ground state and is therefore comparatively insensitive to first-order electronic magnetic and ligand-field fluctuations. In contrast, Er$^{3+}$ is a Kramers ion with a ${}^4I_{15/2}$ ($J=15/2$) ground-state manifold and a large electronic magnetic moment, while Tm$^{3+}$ is non-Kramers but has a ${}^3H_6$ ($J=6$) ground-state manifold. Molecular vibrations can modulate the ligand field and hence the crystal-field and effective spin Hamiltonians, providing spin-lattice relaxation and dephasing channels; low-energy molecular modes have been shown to play a particularly important role in molecular spin relaxation~\cite{moseley2018spinphonon, garlatti2023critical}. Er$^{3+}$ has the additional susceptibility to magnetic spectral diffusion arising from fluctuations of nearby electronic and nuclear spins~\cite{bottger2006optical}. High-energy ligand vibrations may also be important. For Er$^{3+}$, the $\sim6500~\mathrm{cm^{-1}}$ optical energy lies close to the first C--H overtone region ($5800$--$6200~\mathrm{cm^{-1}}$), consistent with the strong C--H-mediated quenching reported for related Er molecular complexes and with the short $T_{1,\mathrm{opt}}\approx0.88~\mu\mathrm{s}$ measured here~\cite{tan2006quenching}. For Tm$^{3+}$, the ${}^3H_4\rightarrow{}^3H_5$ gap ($\sim4348~\mathrm{cm^{-1}}$) overlaps the C--H stretch+bend combination-band region. Although these high-energy vibrations are far above the relevant spin and hyperfine splittings and are therefore unlikely to drive direct spin transitions at cryogenic equilibrium, they still modulate the local electronic Hamiltonian. Moreover, nonradiative relaxation during optical burning can create nonequilibrium vibrational excitation, which may affect the spin/crystal-field environment directly or redistribute anharmonically into lower-energy molecular and lattice modes~\cite{hill1988vibrational,hill1988model}. Such processes could enhance relaxation, dephasing, or spectral diffusion and thereby reduce SHB contrast. Importantly, rapid optical relaxation alone does not establish inefficient shelving, since the latter also depends on decay branching ratios and shelving-state lifetimes, which are not determined here.

Another important outcome of this work is the identification of light-induced local heating as a major effect in these molecular crystals unless proper thermalization is in place. In the Er$^{3+}$ complex, we tracked the population dynamics of the second crystal-field level in the ground-state manifold, $|2\rangle_g$, after optical pumping. By combining the measured optical-depth transient of the $|2\rangle_g \rightarrow |1\rangle_e$ transition with the Boltzmann distribution calculated from the resolved crystal-field energies, we extracted the temperature dynamics of the illuminated region of the sample, as shown in Figure~\ref{fig:Temp_vs_time}. These measurements reveal slow local rethermalization on the timescale of seconds. A similar slow relaxation was also observed in the 100\%-concentration Tm$^{3+}$ complex, where the transmitted probe intensity decays with a characteristic time constant of $2.0 \pm 0.2$~s, indicating that light-induced heating is not unique to Er$^{3+}$ but is a broader feature of these powder-like molecular systems.

We also examined the temperature dependence of the optical absorption linewidth and transition energy for the Er$^{3+}$ complexes at 1\% and 100\% concentration and for the 100\%-concentration Tm$^{3+}$ complex. Both the linewidth broadening and the thermal shift can be described using models based on phonon coupling and an effective energy scale $\Delta E$, as given by Eqs.~\ref{eq:3} and \ref{eq:4}. For the Er$^{3+}$ complexes, the fitted values of $\Delta E$ are consistent with the relevant crystal-field energy separations, while remaining best interpreted as effective energy scales arising from broadened distributions of splittings rather than as exact single-level separations. For the Tm$^{3+}$ complex, the fitted parameters have larger uncertainties, presumably because of the more limited dataset and the stronger variation of the linewidth at elevated temperatures. Since the full crystal-field level structure of the Tm$^{3+}$ complex has not yet been resolved, the fitted $\Delta E$ for Tm$^{3+}$ cannot presently be assigned to a specific crystal-field splitting.

Finally, the concentration-dependent linewidth measurements in the Er$^{3+}$ complexes provide further insight into the origin of the broadening. If the linewidth were dominated mainly by static compositional disorder, one might expect a bell-shaped dependence on concentration, with the largest broadening near an intermediate concentration. Instead, the measured inhomogeneous linewidth increases monotonically from $7.9 \pm 0.3$~GHz at 1\% concentration to $18.1 \pm 0.9$~GHz at 25\% concentration and $40.2 \pm 1.1$~GHz at 100\% concentration. This trend suggests that Er$^{3+}$--Er$^{3+}$ interactions and concentration-dependent dynamical broadening become increasingly important at higher Er$^{3+}$ density. Consistent with this interpretation, no SHB signal was observed even when the readout scanned over a few gigahertz. Therefore, if a spectral hole is formed in the Er$^{3+}$ complex, it must be broadened over a comparable frequency scale and would not appear as a narrow feature within our scan window. Taken together, the monotonic concentration dependence and the absence of a resolvable spectral hole suggest that the homogeneous linewidth may be very large, possibly approaching the same order of magnitude as the measured inhomogeneous linewidth.

Overall, our results indicate that the Eu$^{3+}$ complex is the most promising among the three systems studied for coherent optical spectroscopy. In contrast, the Er$^{3+}$ and Tm$^{3+}$ complexes highlight the importance of suppressing multiphonon relaxation and improving thermalization in powder-like samples. These findings suggest that future studies of molecular powders would benefit from cryogenic platforms with higher cooling power, such as liquid-helium bath or gas-exchange cryostats, which could provide more efficient rethermalization in the locally illuminated region of the sample. Even in the present systems, spectroscopic measurements that require only low optical power may remain feasible. Another promising direction is to study larger, individually mounted single crystals in place of the ensembles of as-grown, needle-like microcrystallites used here, with improved thermal contact to the sample holder potentially enabling better thermalization and reduced thermal broadening. For the Er$^{3+}$ and Tm$^{3+}$ complexes in particular, further progress will likely require molecular designs that reduce multiphonon relaxation sufficiently for coherence properties to become observable. Such advances will be important for assessing the potential of rare-earth molecular materials for quantum photonics and related coherent optical technologies.

\section*{Acknowledgment}
The authors acknowledge the use of OpenAI ChatGPT (GPT-5.6 and GPT-6) during manuscript preparation in 2026 to assist with language editing and the development of supporting Python code for data processing, analysis, and visualization. All AI-assisted outputs were reviewed and verified by the authors, who take full responsibility for the accuracy, scientific content, and integrity of the work.

\section*{Conflicts of Interest}
The authors declare no conflicts of interests.

\section*{Data Availability Statement}
The data that support the findings of this study are available from the corresponding author upon reasonable request

\section*{Supporting Information}
Additional supporting information is provided in a separate Supporting Information document.

\bibliography{sample}

@article{lvovsky2009optical,
  title={Optical quantum memory},
  author={Lvovsky, Alexander I and Sanders, Barry C and Tittel, Wolfgang},
  journal={Nature photonics},
  volume={3},
  number={12},
  pages={706--714},
  year={2009},
  publisher={Nature Publishing Group UK London},
  url = {https://www.nature.com/articles/nphoton.2009.231}
}

@article{heshami2016quantum,
  title={Quantum memories: emerging applications and recent advances},
  author={Heshami, Khabat and England, Duncan G and Humphreys, Peter C and Bustard, Philip J and Acosta, Victor M and Nunn, Joshua and Sussman, Benjamin J},
  journal={Journal of modern optics},
  volume={63},
  number={20},
  pages={2005--2028},
  year={2016},
  publisher={Taylor \& Francis},
  url = {https://www.tandfonline.com/doi/full/10.1080/09500340.2016.1148212}
}

@article{tittel2025quantum,
  title={Quantum networks using rare-earth ions},
  author={Tittel, Wolfgang and Afzelius, Mikael and Kinos, Adam and Rippe, Lars and Walther, Andreas},
  journal={Quantum Science and Technology},
  volume={10},
  number={3},
  pages={033002},
  year={2025},
  publisher={IOP Publishing},
  url = {https://iopscience.iop.org/article/10.1088/2058-9565/addd93/meta}
}

@article{stevenson2022erbium,
  title={Erbium-implanted materials for quantum communication applications},
  author={Stevenson, Paul and Phenicie, Christopher M and Gray, Isaiah and Horvath, Sebastian P and Welinski, Sacha and Ferrenti, Austin M and Ferrier, Alban and Goldner, Philippe and Das, Sujit and Ramesh, Ramamoorthy and others},
  journal={Physical Review B},
  volume={105},
  number={22},
  pages={224106},
  year={2022},
  publisher={APS},
  url = {https://journals.aps.org/prb/abstract/10.1103/PhysRevB.105.224106}
}

@incollection{goldner2015rare,
  title={Rare earth-doped crystals for quantum information processing},
  author={Goldner, Philippe and Ferrier, Alban and Guillot-No{\"e}l, Olivier},
  booktitle={Handbook on the Physics and Chemistry of Rare Earths},
  volume={46},
  pages={1--78},
  year={2015},
  publisher={Elsevier},
  url = {https://www.sciencedirect.com/science/article/abs/pii/B9780444632609002674}
}

@book{liu2006spectroscopic,
  title={Spectroscopic properties of rare earths in optical materials},
  author={Liu, Guokui and Jacquier, Bernard},
  volume={83},
  year={2006},
  publisher={Springer Science \& Business Media},
  url = {https://link.springer.com/book/10.1007/3-540-28209-2}
}

@article{bottger2016optical,
  title={Optical spectroscopy and decoherence studies of Y b 3+: YAG at 968 nm},
  author={B{\"o}ttger, Thomas and Thiel, CW and Cone, RL and Sun, Y and Faraon, A},
  journal={Physical Review B},
  volume={94},
  number={4},
  pages={045134},
  year={2016},
  publisher={APS},
  url = {https://journals.aps.org/prb/abstract/10.1103/PhysRevB.94.045134}
}

@inproceedings{bottger2003material,
  title={Material optimization of Er$^{3+}$:Y$_2$SiO$_5$ at 1.5~$\mu$m for optical processing, memory, and laser frequency stabilization applications},
  author={B{"o}ttger, Thomas and Sun, Yongchen and Thiel, Charles W. and Cone, Rufus L.},
  booktitle={Advanced Optical Data Storage},
  series={Proceedings of SPIE},
  volume={4988},
  pages={51},
  year={2003},
  doi={10.1117/12.474784}
}

@article{young1966apparent,
  title={Apparent lowering of energy levels as measured by Orbach relaxation rates},
  author={Young, B Arlen and Stapleton, HJ},
  journal={Physics Letters},
  volume={21},
  number={5},
  pages={498--501},
  year={1966},
  publisher={Elsevier},
  url = {https://www.sciencedirect.com/science/article/pii/0031916366912716?via%3Dihub}
}

@article{mccumber1963linewidth,
  title={Linewidth and temperature shift of the R lines in ruby},
  author={McCumber, DE and Sturge, MD},
  journal={Journal of Applied Physics},
  volume={34},
  number={6},
  pages={1682--1684},
  year={1963},
  publisher={American Institute of Physics},
  url = {https://pubs.aip.org/aip/jap/article/34/6/1682/362804/Linewidth-and-Temperature-Shift-of-the-R-Lines-in}
}

@article{saglamyurek2015quantum,
  title={Quantum storage of entangled telecom-wavelength photons in an erbium-doped optical fibre},
  author={Saglamyurek, Erhan and Jin, Jeongwan and Verma, Varun B. and Shaw, Matthew D. and Marsili, Francesco and Nam, Sae Woo and Oblak, Daniel and Tittel, Wolfgang},
  journal={Nature Photonics},
  volume={9},
  number={2},
  pages={83--87},
  year={2015},
  doi={10.1038/nphoton.2014.311}
}

@article{kumar2021optical,
  title={Optical spin-state polarization in a binuclear europium complex towards molecule-based coherent light-spin interfaces},
  author={Kumar, Kuppusamy Senthil and Serrano, Diana and Nonat, Aline M. and Heinrich, Beno{\^i}t and Karmazin, Lydia and Charbonni{\`e}re, Lo{\"i}c J. and Goldner, Philippe and Ruben, Mario},
  journal={Nature Communications},
  volume={12},
  pages={2152},
  year={2021},
  doi={10.1038/s41467-021-22383-x}
}

@article{serrano2022ultranarrow,
  title={Ultra-narrow optical linewidths in rare-earth molecular crystals},
  author={Serrano, Diana and Kuppusamy, Senthil Kumar and Heinrich, Beno{\^i}t and Fuhr, Olaf and Hunger, David and Goldner, Philippe and Ruben, Mario},
  journal={Nature},
  volume={603},
  pages={241--246},
  year={2022},
  doi={10.1038/s41586-021-04316-2}
}

@article{rasekh2026ultra,
  title={Ultra-narrow homogeneous linewidths of erbium-doped silica glass fibers at millikelvin temperatures: magnetic field and temperature dependence},
  author={Rasekh, Farhad and Kamel, Nasser Gohari and Bornadel, Mahdi and Kumar, Sourabh and Saglamyurek, Erhan and Simon, Christoph and Oblak, Daniel},
  journal={Optica Quantum},
  volume={4},
  number={1},
  pages={63--72},
  year={2026},
  publisher={Optica Publishing Group},
  url = {https://opg.optica.org/opticaq/fulltext.cfm?uri=opticaq-4-1-63}
}

@article{kuppusamy2023observation,
  title={Observation of narrow optical homogeneous linewidth and long nuclear spin lifetimes in a prototypical [Eu (trensal)] complex},
  author={Kuppusamy, Senthil Kumar and Vasilenko, Evgenij and Li, Weizhe and Hessenauer, Jannis and Ioannou, Christina and Fuhr, Olaf and Hunger, David and Ruben, Mario},
  journal={The Journal of Physical Chemistry C},
  volume={127},
  number={22},
  pages={10670--10679},
  year={2023},
  publisher={ACS Publications},
  url = {https://pubs.acs.org/doi/full/10.1021/acs.jpcc.3c02903}
}

@article{zhong2015nanophotonic,
  title={Nanophotonic coherent light--matter interfaces based on rare-earth-doped crystals},
  author={Zhong, Tian and Kindem, Jonathan M. and Miyazono, Evan and Faraon, Andrei},
  journal={Nature Communications},
  volume={6},
  pages={8206},
  year={2015},
  doi={10.1038/ncomms9206},
  url = {https://www.nature.com/articles/ncomms9206}
}

@article{kolesov2012optical,
  title={Optical detection of a single rare-earth ion in a crystal},
  author={Kolesov, R. and Xia, K. and Reuter, R. and St\"ohr, R. and Zappe, A. and Meijer, J. and Hemmer, P. R. and Wrachtrup, J.},
  journal={Nature Communications},
  volume={3},
  pages={1029},
  year={2012},
  doi={10.1038/ncomms2034},
  url = {https://www.nature.com/articles/ncomms2034}
}

@article{vasilenko2026optically,
  title={Optically detected nuclear magnetic resonance of coherent spins in a molecular complex},
  author={Vasilenko, Evgenij and Unni Chorakkunnath, Vishnu and Resch, Jeremias and Jobbitt, Nicholas and Serrano, Diana and Goldner, Philippe and Kuppusamy, Senthil Kumar and Ruben, Mario and Hunger, David},
  journal={Nature Materials},
  pages={1--6},
  year={2026},
  publisher={Nature Publishing Group UK London},
  url = {https://www.nature.com/articles/s41563-026-02539-0}
}

@article{vasilenko2026controlled,
  title={Controlled ion-ion interactions and cavity-enhanced emission of a coherent dinuclear Eu$^{3+}$ complex},
  author={Vasilenko, Evgenij and Unni Chorakkunnath, Vishnu and Brachnakova, Barbora and Jobbitt, Nicholas Lester and Kuppusamy, Senthil Kumar and Hunger, David and Ruben, Mario},
  journal={arXiv preprint arXiv:2606.11947},
  year={2026},
  doi={10.48550/arXiv.2606.11947}
}

@article{bornadel2025hole,
  title={Hole-burning experiments and modeling in erbium-doped silica glass fibers down to millikelvin temperatures: Evidence for ultralong population storage},
  author={Bornadel, Mahdi and Shafiei Alavijeh, Sara and Rasekh, Farhad and Gohari Kamel, Nasser and Kimiaee Asadi, Faezeh and Saglamyurek, Erhan and Oblak, Daniel and Simon, Christoph},
  journal={Physical Review Applied},
  volume={24},
  number={1},
  pages={014030},
  year={2025},
  publisher={APS},
  url = {https://journals.aps.org/prapplied/abstract/10.1103/dtns-snw7}
}

@article{wang2025nuclear,
  title={Nuclear spins in a solid exceeding 10-hour coherence times for ultra-long-term quantum storage},
  author={Wang, Fudong and Ren, Miaomiao and Sun, Weiye and Guo, Mucheng and Sellars, Matthew J and Ahlefeldt, Rose L and Bartholomew, John G and Yao, Juan and Liu, Shuping and Zhong, Manjin},
  journal={PRX Quantum},
  volume={6},
  number={1},
  pages={010302},
  year={2025},
  publisher={APS},
  url = {https://journals.aps.org/prxquantum/abstract/10.1103/PRXQuantum.6.010302}
}

@article{le2021twenty,
  title={Twenty-three--millisecond electron spin coherence of erbium ions in a natural-abundance crystal},
  author={Le Dantec, Marianne and Ran{\v{c}}i{\'c}, Milo{\v{s}} and Lin, Sen and Billaud, Eric and Ranjan, Vishal and Flanigan, Daniel and Bertaina, Sylvain and Chaneli{\`e}re, Thierry and Goldner, Philippe and Erb, Andreas and others},
  journal={Science advances},
  volume={7},
  number={51},
  pages={eabj9786},
  year={2021},
  publisher={American Association for the Advancement of Science},
  url = {https://www.science.org/doi/10.1126/sciadv.abj9786}
}

@article{bayliss2020optically,
  title={Optically addressable molecular spins for quantum information processing},
  author={Bayliss, SL and Laorenza, DW and Mintun, PJ and Kovos, BD and Freedman, DE and Awschalom, DD},
  journal={Science},
  volume={370},
  number={6522},
  pages={1309--1312},
  year={2020},
  publisher={American Association for the Advancement of Science},
  url = {https://www.science.org/doi/full/10.1126/science.abb9352}
}

@article{saglamyurek2016multiplexed,
  title={A multiplexed light-matter interface for fibre-based quantum networks},
  author={Saglamyurek, Erhan and Grimau Puigibert, Marcelli and Zhou, Qiang and Giner, Lambert and Marsili, Francesco and Verma, Varun B and Woo Nam, Sae and Oesterling, Lee and Nippa, David and Oblak, Daniel and others},
  journal={Nature communications},
  volume={7},
  number={1},
  pages={11202},
  year={2016},
  publisher={Nature Publishing Group UK London},
  url = {https://www.nature.com/articles/ncomms11202}
}

@article{jin2015telecom,
  title={Telecom-wavelength atomic quantum memory in optical fiber for heralded polarization qubits},
  author={Jin, Jeongwan and Saglamyurek, Erhan and Puigibert, Marcel l{\'\i} Grimau and Verma, Varun and Marsili, Francesco and Nam, Sae Woo and Oblak, Daniel and Tittel, Wolfgang},
  journal={Physical review letters},
  volume={115},
  number={14},
  pages={140501},
  year={2015},
  publisher={APS},
  url = {https://journals.aps.org/prl/abstract/10.1103/PhysRevLett.115.140501}
}

@article{saglamyurek2011broadband,
  title={Broadband waveguide quantum memory for entangled photons},
  author={Saglamyurek, Erhan and Sinclair, Neil and Jin, Jeongwan and Slater, Joshua A and Oblak, Daniel and Bussieres, F{\'e}lix and George, Mathew and Ricken, Raimund and Sohler, Wolfgang and Tittel, Wolfgang},
  journal={Nature},
  volume={469},
  number={7331},
  pages={512--515},
  year={2011},
  publisher={Nature Publishing Group UK London},
  url = {https://www.nature.com/articles/nature09719}
}

@article{davidson2020improved,
  title={Improved light-matter interaction for storage of quantum states of light in a thulium-doped crystal cavity},
  author={Davidson, Jacob H and Lefebvre, Pascal and Zhang, Jun and Oblak, Daniel and Tittel, Wolfgang},
  journal={Physical Review A},
  volume={101},
  number={4},
  pages={042333},
  year={2020},
  publisher={APS},
  url = {https://journals.aps.org/pra/abstract/10.1103/PhysRevA.101.042333}
}

@article{hastings2008spectral,
  title={Spectral hole-burning spectroscopy in Nd 3+: YVO 4},
  author={Hastings-Simon, SR and Afzelius, M and Min{\'a}{\v{r}}, J and Staudt, MU and Lauritzen, B and De Riedmatten, H and Gisin, N and Amari, Atia and Walther, Andreas and Kr{\"o}ll, Stefan and others},
  journal={Physical Review B—Condensed Matter and Materials Physics},
  volume={77},
  number={12},
  pages={125111},
  year={2008},
  publisher={APS},
  url = {https://journals.aps.org/prb/abstract/10.1103/PhysRevB.77.125111}
}

@article{konz2003temperature,
  title={Temperature and concentration dependence of optical dephasing, spectral-hole lifetime, and anisotropic absorption in Eu 3+: Y 2 SiO 5},
  author={K{\"o}nz, Flurin and Sun, Y and Thiel, CW and Cone, RL and Equall, RW and Hutcheson, RL and Macfarlane, RM},
  journal={Physical Review B},
  volume={68},
  number={8},
  pages={085109},
  year={2003},
  publisher={APS},
  url = {https://journals.aps.org/prb/abstract/10.1103/PhysRevB.68.085109}
}

@article{macfarlane1987spectral,
  title={Spectral hole burning by population storage in Zeeman sublevels of LaFsub3: Nd 3+},
  author={Macfarlane, RM and Vial, JC},
  journal={Physical Review B},
  volume={36},
  number={7},
  pages={3511},
  year={1987},
  publisher={APS},
  url = {https://journals.aps.org/prb/abstract/10.1103/PhysRevB.36.3511}
}

@article{swinehart1962beer,
  title={The beer-lambert law},
  author={Swinehart, Donald F},
  journal={Journal of chemical education},
  volume={39},
  number={7},
  pages={333},
  year={1962},
  publisher={ACS Publications},
  url = {https://pubs.acs.org/doi/10.1021/ed039p333}
}

@article{kunkel2016dephasing,
  title={Dephasing mechanisms of optical transitions in rare-earth-doped transparent ceramics},
  author={Kunkel, Nathalie and Bartholomew, John and Welinski, Sacha and Ferrier, Alban and Ikesue, Akio and Goldner, Philippe},
  journal={Physical Review B},
  volume={94},
  number={18},
  pages={184301},
  year={2016},
  publisher={APS},
  url = {https://journals.aps.org/prb/abstract/10.1103/PhysRevB.94.184301}
}

@article{liu2020defect,
  title={Defect engineering for quantum grade rare-earth nanocrystals},
  author={Liu, Shuping and Fossati, Alexandre and Serrano, Diana and Tallaire, Alexandre and Ferrier, Alban and Goldner, Philippe},
  journal={ACS nano},
  volume={14},
  number={8},
  pages={9953--9962},
  year={2020},
  publisher={ACS Publications},
  url = {https://pubs.acs.org/doi/full/10.1021/acsnano.0c02971}
}

@article{ahlefeldt2013optical,
  title={Optical lifetime and linewidth studies of the 7F0→ 5D0 transition in EuCl3{\textperiodcentered} 6H2O: A potential material for quantum memory applications},
  author={Ahlefeldt, RL and Manson, NB and Sellars, MJ},
  journal={Journal of luminescence},
  volume={133},
  pages={152--156},
  year={2013},
  publisher={Elsevier},
  url = {https://www.sciencedirect.com/science/article/pii/S0022231311006892}
}

@article{galico2025magneto,
  title={Magneto-Luminescence Thermometry with Magnetic Circularly Polarized Luminescence},
  author={G{\'a}lico, Diogo Alves and Murugesu, Muralee},
  journal={Angewandte Chemie International Edition},
  volume={64},
  number={35},
  pages={e202505806},
  year={2025},
  publisher={Wiley Online Library},
  url = {https://onlinelibrary.wiley.com/doi/full/10.1002/anie.202505806}
}

@article{teixeira2016thermal,
  title={Thermal and spectroscopic study to investigate p-aminobenzoic acid, sodium p-aminobenzoate and its compounds with some lighter trivalent lanthanides},
  author={Teixeira, JA and Nunes, WDG and Colman, TAD and Do Nascimento, ALCS and Caires, FJ and Campos, FX and G{\'a}lico, DA and Ionashiro, M},
  journal={Thermochimica acta},
  volume={624},
  pages={59--68},
  year={2016},
  publisher={Elsevier},
  url = {https://www.sciencedirect.com/science/article/pii/S0040603115004748}
}

@article{neelgund2007tris,
  title={Tris (acetylacetonato-$\kappa$2O, O$\prime$)(1, 10-phenanthroline-$\kappa$2N, N$\prime$) erbium (III)},
  author={Neelgund, Gururaj M and Shivashankar, SA and Narasimhamurthy, T and Rathore, RS},
  journal={Crystal Structure Communications},
  volume={63},
  number={2},
  pages={m74--m76},
  year={2007},
  publisher={International Union of Crystallography},
  url = {https://journals.iucr.org/paper?S0108270107000248}
}

@article{moseley2018spinphonon,
  title={Spin--phonon couplings in transition metal complexes with slow magnetic relaxation},
  author={Moseley, Duncan H. and Stavretis, Shelby E. and Thirunavukkuarasu, Komalavalli and Ozerov, Mykhaylo and Cheng, Yongqiang and Daemen, Luke L. and Ludwig, Jonathan and Lu, Zhengguang and Smirnov, Dmitry and Brown, Craig M. and Pandey, Anup and Ramirez-Cuesta, A. J. and Lamb, Adam C. and Atanasov, Mihail and Bill, Eckhard and Neese, Frank and Xue, Zi-Ling},
  journal={Nature Communications},
  volume={9},
  pages={2572},
  year={2018},
  doi={10.1038/s41467-018-04896-0}
}

@article{garlatti2023critical,
  title={The critical role of ultra-low-energy vibrations in the relaxation dynamics of molecular qubits},
  author={Garlatti, Elena and Albino, Andrea and Chicco, Simone and Nguyen, V. H. A. and Santanni, Fabio and Paolasini, Luigi and Mazzoli, Claudio and Caciuffo, Roberto and Totti, Federico and Santini, Paolo and Sessoli, Roberta and Lunghi, Alessandro and Carretta, Stefano},
  journal={Nature Communications},
  volume={14},
  pages={1653},
  year={2023},
  doi={10.1038/s41467-023-36852-y}
}

@article{tan2006quenching,
  title={Quenching of IR Luminescence of Erbium, Neodymium, and Ytterbium {$\beta$}-Diketonate Complexes by Ligand C--H and C--D Bonds},
  author={Tan, Rendy H. C. and Motevalli, Majid and Abrahams, Isaac and Wyatt, Peter B. and Gillin, William P.},
  journal={The Journal of Physical Chemistry B},
  volume={110},
  number={48},
  pages={24476--24479},
  year={2006},
  doi={10.1021/jp0654462}
}

@article{hill1988vibrational,
  title={Vibrational relaxation and vibrational cooling in low temperature molecular crystals},
  author={Hill, Jeffrey R. and Chronister, Eric L. and Chang, Ta-Chau and Kim, Hackjin and Postlewaite, Jay C. and Dlott, Dana D.},
  journal={The Journal of Chemical Physics},
  volume={88},
  number={2},
  pages={949--967},
  year={1988},
  doi={10.1063/1.454175}
}

@article{hill1988model,
  title={A model for ultrafast vibrational cooling in molecular crystals},
  author={Hill, Jeffrey R. and Dlott, Dana D.},
  journal={The Journal of Chemical Physics},
  volume={89},
  number={2},
  pages={830--841},
  year={1988},
  doi={10.1063/1.455206}
}

@article{bottger2006optical,
  title={Optical decoherence and spectral diffusion at 1.5~{$\mu$}m in Er$^{3+}$:Y$_2$SiO$_5$ versus magnetic field, temperature, and Er$^{3+}$ concentration},
  author={B{\"o}ttger, Thomas and Thiel, C. W. and Sun, Y. and Cone, R. L.},
  journal={Physical Review B},
  volume={73},
  pages={075101},
  year={2006},
  doi={10.1103/PhysRevB.73.075101}
}

@article{garlatti2020unveiling,
  title={Unveiling phonons in a molecular qubit with four-dimensional inelastic neutron scattering and density functional theory},
  author={Garlatti, E and Tesi, L and Lunghi, A and Atzori, M and Voneshen, DJ and Santini, P and Sanvito, S and Guidi, T and Sessoli, R and Carretta, S},
  journal={Nature communications},
  volume={11},
  number={1},
  pages={1751},
  year={2020},
  publisher={Nature Publishing Group UK London},
  url = {https://www.nature.com/articles/s41467-020-15475-7}
}

@article{lu2019abinitio,
  title={Ab initio molecular dynamics simulation of vibrational energy redistribution of selective excitation of C--H stretching vibrations for solid nitromethane},
  author={Lu, Meilin and Zheng, Zhaoyang and Zhu, Gangbei and Wang, Yuxiao and Yang, Yanqiang},
  journal={Physical Chemistry Chemical Physics},
  volume={21},
  number={37},
  pages={20822--20828},
  year={2019},
  doi={10.1039/C9CP03797G}
}

@article{schwartz1974electron,
  title={Electron paramagnetic resonance study of {Ce$^{3+}$}, {Dy$^{3+}$} and {Yb$^{3+}$} in {Cs$_2$NaYCl$_6$}. A crystal with sites of perfect octahedral symmetry},
  author={Schwartz, Robert W. and Hill, Nicholas J.},
  journal={Journal of the Chemical Society, Faraday Transactions 2: Molecular and Chemical Physics},
  volume={70},
  pages={124--131},
  year={1974},
  doi={10.1039/F29747000124}
}

@article{guillot2006hyperfine,
  title={Hyperfine interaction of Er 3+ ions in Y 2 Si O 5: An electron paramagnetic resonance spectroscopy study},
  author={Guillot-No{\"e}l, O and Goldner, Ph and Du, Y Le and Baldit, E and Monnier, P and Bencheikh, K},
  journal={Physical Review B—Condensed Matter and Materials Physics},
  volume={74},
  number={21},
  pages={214409},
  year={2006},
  publisher={APS},
  url = {https://journals.aps.org/prb/abstract/10.1103/PhysRevB.74.214409}
}

\end{document}